\documentclass[%
 aip,
 amsmath,amssymb,
preprint,%
]{revtex4-1}

\usepackage{graphicx}
\usepackage{dcolumn}
\usepackage{bm}
\usepackage{braket}
\usepackage{physics}
\usepackage[utf8]{inputenc}
\usepackage[T1]{fontenc}
\usepackage{mathptmx}
\usepackage[font={small, normalfont}, textfont=rm, labelfont={bf}, labelsep=colon]{caption} 
\usepackage{subcaption}
\usepackage{setspace}
\usepackage[margin=3cm, top=2.5cm, bottom=3cm]{geometry}
\usepackage[dvipsnames]{xcolor}
\usepackage[section]{placeins}
\usepackage{xparse,xcoffins}
\usepackage[normalem]{ulem}
\usepackage{booktabs}

\usepackage{array}
\newcolumntype{L}[1]{>{\raggedright\let\newline\\\arraybackslash\hspace{0pt}}m{#1}}
\newcolumntype{C}[1]{>{\centering\let\newline\\\arraybackslash\hspace{0pt}}m{#1}}
\newcolumntype{R}[1]{>{\raggedleft\let\newline\\\arraybackslash\hspace{0pt}}m{#1}}

\newcommand{\agtwo}{$2^1A_g^-$}
\newcommand{\agthree}{$3^1A_g^-$}
\newcommand{\bum}{$1^1B_u^-$}
\newcommand{\bu}{$1^1B_u^+$}
\newcommand{\triplet}{$1^3B_u^-$}

\newcommand{\Q}{$1^5A_g^-$}

\begin{document}

\ExplSyntaxOn
\NewCoffin\imagecoffin
\NewCoffin\labelcoffin

\keys_define:nn { miguel/label }
 {
  label   .tl_set:N = \l_miguel_label_tl,
  labelbox .bool_set:N = \l_miguel_label_box_bool,
  labelbox .default:n = true,
  fontsize .tl_set:N = \l_miguel_label_size_tl,
  fontsize .initial:n = \footnotesize,
  pos .choice:,
  pos/nw .code:n = \tl_set:Nn \l_miguel_label_pos_tl { left,up },
  pos/ne .code:n = \tl_set:Nn \l_miguel_label_pos_tl { right,up },
  pos/sw .code:n = \tl_set:Nn \l_miguel_label_pos_tl { left,down },
  pos/se .code:n = \tl_set:Nn \l_miguel_label_pos_tl { right,down },
  pos/n .code:n = \tl_set:Nn \l_miguel_label_pos_tl { hc,up },
  pos/w .code:n = \tl_set:Nn \l_miguel_label_pos_tl { left,vc },
  pos/s .code:n = \tl_set:Nn \l_miguel_label_pos_tl { hc,down },
  pos/e .code:n = \tl_set:Nn \l_miguel_label_pos_tl { right,vc },
  pos .initial:n = nw,
  unknown .code:n   = \clist_put_right:Nx \l_miguel_label_clist
                       { \l_keys_key_tl = \exp_not:n { #1 } }
 }
\clist_new:N \l_miguel_label_clist
\box_new:N \l_miguel_label_box
\box_new:N \l_miguel_label_image_box

\NewDocumentCommand{\xincludegraphics}{O{}m}
 {
  \group_begin:
  \tl_clear:N \l_miguel_label_tl
  \clist_clear:N \l_miguel_label_clist
  \keys_set:nn { miguel/label } { #1 }
  \tl_if_empty:NTF \l_miguel_label_tl
   {
    \miguel_includegraphics:Vn \l_miguel_label_clist { #2 }
   }
   {
    \SetHorizontalCoffin\imagecoffin
     {
      \miguel_includegraphics:Vn \l_miguel_label_clist { #2 }
     }
    \SetHorizontalCoffin\labelcoffin
     {
      \raisebox{\depth}
       {
        \bool_if:NTF \l_miguel_label_box_bool
         { \fcolorbox{white}{white}{\l_miguel_label_size_tl\l_miguel_label_tl} }
         { \l_miguel_label_size_tl\l_miguel_label_tl }
       }
     }
    \SetVerticalPole\imagecoffin{left}{3pt+\CoffinWidth\labelcoffin/2}
    \SetVerticalPole\imagecoffin{right}{\Width-3pt-\CoffinWidth\labelcoffin/2}
    \SetHorizontalPole\imagecoffin{up}{\Height-3pt-\CoffinHeight\labelcoffin/2}
    \SetHorizontalPole\imagecoffin{down}{3pt+\CoffinHeight\labelcoffin/2}
    \use:x{\JoinCoffins\imagecoffin[\l_miguel_label_pos_tl]\labelcoffin[vc,hc]}
    \TypesetCoffin\imagecoffin
   }
   \group_end:
 }
\NewDocumentCommand{\setlabel}{m}
 {
  \keys_set:nn { miguel/label } { #1 }
 }

\cs_new_protected:Nn \miguel_includegraphics:nn
 {
  \includegraphics[#1]{#2}
 }
\cs_generate_variant:Nn \miguel_includegraphics:nn { V }

\ExplSyntaxOff


\title[Higher energy triplet-pair states in polyenes and their role in intramolecular singlet fission]{Higher energy triplet-pair states in polyenes and their role in intramolecular singlet fission}

\author{D. J. Valentine}
\email{darren.valentine@chem.ox.ac.uk}
\affiliation{Department of Chemistry, Physical and Theoretical Chemistry Laboratory, University of Oxford, Oxford, OX1 3QZ, United Kingdom
}%
\affiliation{Balliol College, University of Oxford, Oxford, OX1 3BJ, United Kingdom}
\author{D. Manawadu}
\affiliation{Department of Chemistry, Physical and Theoretical Chemistry Laboratory, University of Oxford, Oxford, OX1 3QZ, United Kingdom
}
\author{W. Barford}
\email{william.barford@chem.ox.ac.uk}%
\affiliation{Department of Chemistry, Physical and Theoretical Chemistry Laboratory, University of Oxford, Oxford, OX1 3QZ, United Kingdom
}%
\affiliation{Balliol College, University of Oxford, Oxford, OX1 3BJ, United Kingdom
}%



\begin{abstract}
Probing extended polyene systems with energy in excess of the bright state (\bu / $S_2$) band edge generates triplets via singlet fission. This process is not thought to involve the \agtwo{} / $S_1$ state, suggesting that other states  play a role. Using density matrix renormalisation group (DMRG) calculations of the Pariser-Parr-Pople-Peierls Hamiltonian, we investigate candidate states that could be involved in singlet fission. We find that the relaxed  \bum{} and \agthree{}  singlet states and $1^5A_g^-$ quintet state lie below the $S_2$ state. The   \bum{}, \agthree{} and $1^5A_g^-$ states are all thought to have triplet-triplet character, which is confirmed by our calculations of bond dimerization, spin-spin correlation and wavefunction overlap with products of triplet states.
We thus show that there is a family of singlet excitations (i.e., \agtwo{}, \bum{}, \agthree{}, $\cdots$), composed of both triplet-pair and electron-hole character, which are fundamentally the same excitation, but have different center-of-mass energies. The lowest energy member of this family, the $2^1A_g^-$ state, cannot undergo singlet fission. But higher energy members (e.g., the \agthree{}) state, owing to their increased kinetic energy and reduced electron-lattice relaxation, can undergo singlet fission for certain chain lengths.
\end{abstract}

\maketitle



\section{\label{sec:intro}Introduction}
Current commercially available solar cell technology is impeded by the Shockley–Queisser limit, which means that higher energy photons are not efficiently utilized for electricity generation \cite{Shockley1961}. When a high energy photon is absorbed, the energy greater than the device's band gap is lost as heat. There are a number of different ways to better utilize the solar spectrum, one such option is singlet fission.

Singlet fission is a process in which a singlet exciton generated by photoexcitation evolves into two separate triplets \cite{Smith2013b}.
Many polyene systems have been shown to exhibit this phenomenon  \cite{Lanzani1999,Lanzani2001,Antognazza2010,Musser2013,Kasai2015,Busby2015,Kraabel1998,Musser2019,Huynh2018,Huynh2017,Hu2018}. If the two separate triplets have energy greater than or equal to the band gap of a photovoltaic carrier material, they can each generate a free electron-hole pair in the carrier material \cite{Rao2017}.

Singlet fission is often assumed to involve three processes or steps.
The first step is state interconversion from the initial photoexcited state to a singlet state with triplet-pair character.
After state interconversion the triplet-pair is coherent and correlated.
In the next step, the triplets migrate away from one another; this process can be described as a loss of electronic interaction.
During this step, which is spin allowed, the triplets retain their spin coherence forming a geminate triplet-pair in an overall singlet state.
The final step involves the loss of spin coherence and leads to two independent triplets (or a non-geminate triplet-pair)
 \cite{Marcus2020}.
This step is not spin allowed and is expected to be slower than the preceding steps. Marcus and Barford have recently investigated this step using a Heisenberg spin chain model.
They show how spin orbital coupling and dephasing from the environment determines this process \cite{Marcus2020}.

Polyene systems are often modelled as having $C_{2h}$ symmetry \cite{Bursill1999,Barford2013a,Barford2001,Ren2017,Aryanpour2015,Aryanpour2015a,Schmidt2012a,Tavan1987,Hu2015}.
In this framework, the first excited singlet state, $S_1$, has the same symmetry as the ground state, $1^1A_g^-$, and is therefore optically inactive.
The strongly optically absorbing singlet state is the  $1^1B_u^+$ state. Although this is not generally the second excited singlet state in polyenes,  it is typically labelled $S_2$.
In polyenes some low energy excited states have multiple triplet excitation character\cite{Tavan1987}.
 This is the case for the $2^1A_g^-$ state, which is sometimes considered as a bound pair of triplet excitations \cite{Bursill1999,Barford2013a,Barford2001,Ren2017,Aryanpour2015,Aryanpour2015a,Schmidt2012a,Tavan1987}.

In polyene systems it remains unclear if singlet fission proceeds via the \agtwo \ state, a vibrationally hot variant of the \agtwo \ state, or a  different state \cite{Musser2013, Musser2019, Antognazza2010}. It is also unclear whether singlet fission in polyene type materials is an inter- or  intra-molecular process \cite{Musser2013,Wang2011,Yu2017a}.

It has been observed that in long isolated chains no singlet fission occurs after photoexcitation \textit{at} the band edge. Instead, the system relaxes non-radiatively via the \agtwo \ state. The $S_2$ to $S_1$ transition occurs via internal conversion between the two potential energy surfaces \cite{Taffet2019}, taking place on a time-scale of 100s fs\cite{Musser2013,Antognazza2010,Musser2019}.

\begin{figure}[h]
	\includegraphics[width=\linewidth]{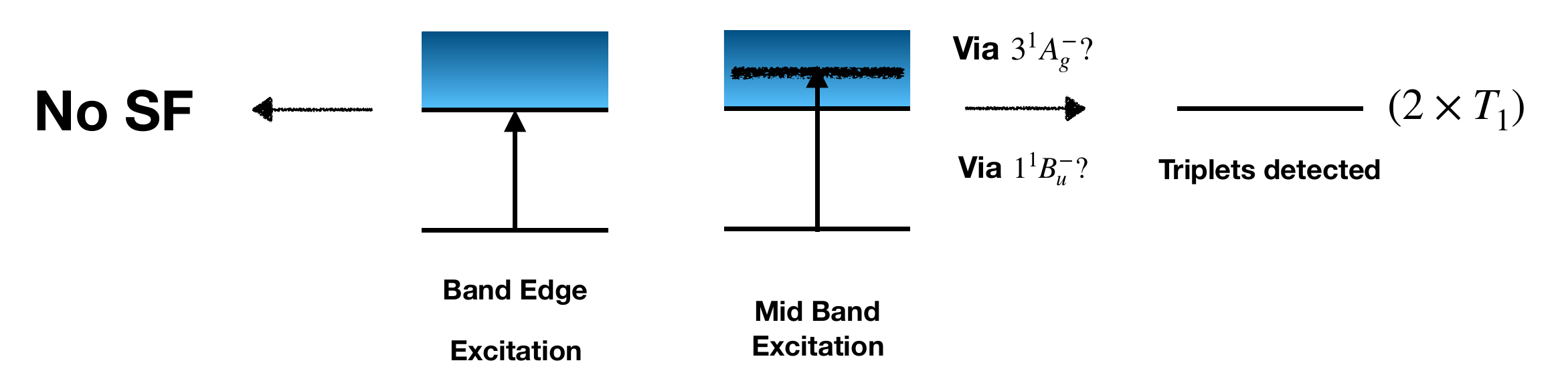}
	\caption{Schematic of potential relaxation pathways from the bright state in polyenes}
	\label{fig:Scheme}
\end{figure}

Upon excitation with energy in excess of the band edge, however, triplets are detected, with isolated triplet signatures appearing in transient absorption spectroscopy measurements \cite{Musser2013,Antognazza2010}.
The occurrence of these triplet signals is attributed to singlet fission. Experiments suggest that this mid-band excited singlet fission does not proceed via the \agtwo \ state.
It is claimed that there are two relaxation pathways: one to the ground state (which proceeds via the \agtwo  \ state) and a different singlet fission pathway with no \agtwo \ involvement, \cite{Musser2013, Antognazza2010} as illustrated in Fig.\ \ref{fig:Scheme}.

If singlet fission in polyenes does not involve the \agtwo \ state, but does require excess energy to overcome a barrier, it can be asked do any higher energy states contribute? Upon vertical excitation it has been found that the $1^1B_u^-$ and $3^1A_g^-$ states exist above the $1^1B_u^+$, although as the chain length increases the $1B_u^-$ energy falls below the $1^1B_u^+$  \cite{Tavan1987,Hashimoto2018}. It is also thought that the $1^1B_u^-$ and $3^1A_g^-$ states have triplet-triplet character \cite{Tavan1987}.

In addition to the singlet triplet-pair state, a quintet triplet-pair  fission intermediate, $^5(T_1 T_1)$ has been observed in acene  materials \cite{Tayebjee2017, Weiss2017a, Sanders2019}. Spin mixing is possible between the $^1(T_1T_1)$ and $^5(T_1T_1)$ states \cite{Merrifield1971}, meaning the quintet could be involved in the singlet fission process or offer an alternative relaxation pathway for the excited molecule.

In this paper we present our calculations of the properties of the key excited states of polyenes, i.e., the \agtwo, \bu, $1^1B_u^-$ and $3^1A_g^-$ singlet states, the \Q{} quintet state and the \triplet{} triplet state.
We use the Parsier-Parr-Pople-Peierls model to describe interacting $\pi$-electrons coupled to the nuclei, which is solved using the density matrix renormalization group (DMRG) method.
We investigate the relaxed geometries of these states within a soliton framework.
Excitations in polyene systems contain spin-density wave, bond-order excitations and charge density waves.
The interplay between these contributions leads to a myriad of  phenomena \citep{Barford2013a}.
To gain insight into the nature of the higher energy excited states, we characterize the states using the spin-spin correlation function, and triplet-pair and electron-hole projections.
We also investigate the optical transitions from these key states.

As we explain in the Discussion Section, we postulate that the \agthree \ state (or another member of the `$2A_g$ family') is the spin-correlated $^1(T \cdots T)$ state, some times referred to as the geminate triplet-pair, observed in the SF process in polyenes.

The paper is organized as follows. In Sec.\ \ref{sec:PPPP} we introduce the Pariser-Parr-Pople-Peierls model. In Sec.\ \ref{sec:energies} we discuss the results of our vertical and relaxed energy calculations.
Sec.\ \ref{sec:bond_dime} discusses the relaxed geometries of the excited states.
In Sec.\ \ref{sec:spin} we use the spin-spin correlation function to characterise the soliton structure of the states.  In Section \ref{sec:wf} we also characterise the states via their electron-hole wavefunctions and their overlap with products of triplet states. In Section \ref{sec:spec} we relate our work to experimental results via a calculation of the excited state spectra and  conclude in Section \ref{sec:con}.

\section{\label{sec:PPPP} Pariser-Parr-Pople-Peierls Model}

We use the Pariser-Parr-Pople-Peierls (PPPP) model to treat the $\pi$-electrons of the conjugated system. This model includes both long-range electronic interactions and electron-nuclear coupling. It is defined as\cite{Barford2013a}
\begin{equation}
	H_{PPPP} = H_{PPP} + H_{el-ph} + H_{elastic},
\end{equation}
where $H_{PPP}$ is the Pariser-Parr-Pople Hamiltonian, defined by
\begin{equation}
H_{PPP} = -2t_0 \sum_n  \hat{T}_n + U \sum_n  \big( N_{n \uparrow} - \frac{1}{2} \big) \big( N_{n \downarrow} - \frac{1}{2} \big)     + \frac{1}{2} \sum_{n \neq m} V_{n, m} \big( N_n -1 \big) \big( N_m -1 \big).
\end{equation}
Here, $\hat{T}_ n = \frac{1}{2}\sum_{\sigma} \left( c^{\dag}_{n , \sigma} c_{n + 1 , \sigma} + c^{\dag}_{n+ 1 , \sigma} c_{n  , \sigma} \right)$ is the bond order operator, $t_0$ is the hopping integral for a uniform, undistorted chain, $U$ is the Coulombic interaction of two electrons in the same orbital and $V_{nm}$ is the long range Coulombic repulsion. We use the Ohno potential given by $V_{nm} =  U / \sqrt{ 1 + (U \epsilon_r r_{nm} / 14.397 )^2 }$, with bond lengths $r_{nm}$ in \AA.

$H_{el-ph}$ is the electron-phonon coupling, given by
\begin{equation}
	H_{el-ph} =  2 \alpha \sum_n  \left( u_{n+1} - u_n \right) \hat{T}_n - 2 \alpha W \sum_n  \left( u_{n+1} - u_n \right) \left( N_{n+1} -1 \right) \left( N_n - 1 \right),
\end{equation}
where $\alpha$ is the electron-nuclear coupling parameter and $u_n$ is the displacement of nucleus $n$ from its undistorted position. Through this term changes in bond length cause changes in the hopping integrals and the Coulomb interactions. Due to the rapid decay of the density-density correlator, $(N_n -1 )(N_m - 1)$, with distance only changes in the Coulomb potential to first order are considered. Therefore, $W$ is determined by
\begin{equation}
	W = \frac{U\gamma r_0}{\left( 1+\gamma r_0^2 \right)^{3/2}},
\end{equation}
where $\gamma = \left( U \epsilon_r / 14.397 \right)^2$ and $r_0$ is the undistorted average bond length in \AA.

The elastic energy of the nuclei contributes through $H_{ph}$, defined as
\begin{equation}
	H_{ph} = \frac{\alpha^2}{\pi t_0 \lambda} \sum_n \left( u_{n+1} - u_n \right)^2 + \Gamma \sum_n \left( u_{n+1} - u_n \right),
\end{equation}
where $\lambda $, the dimensionless electron-nuclear coupling parameter, is $\frac{2\alpha^2}{\pi K t}$ and  $K$ is the nuclear spring constant. $\Gamma$ is a Lagrange multiplier which ensures a constant chain length.

The requirement that the force per bond vanishes at equilibrium gives a self-consistent equation for the bond distortion, namely
\begin{equation}
	  \left( u_{n+1} - u_n \right)  = \frac{ \pi t \lambda }{\alpha} \left(\Gamma - \Braket{ \hat{T_n} }  + W \Braket{ \hat{D_n} } \right),
\end{equation}
where $\hat{D_n} = \left( N_n -1\right) \left( N_{n+1} -1 \right )$ is the nearest neighbor density-density correlator.
We follow a parameterization of the PPP Hamiltonian by Mazumdar and Chandross for \textit{screened} polyacetylene, namely $U=8\ eV$, $\epsilon_r=2$ and $t_0 = 2.4 \ eV$  \cite{Chandross1997}.  We use the  electron-nuclear coupling constants of Barford and co-workers, namely, $\lambda=0.115$ and $\alpha=0.4593 \ eV$ \AA$^{-1} $ \  \cite{Barford2001}.


\section{\label{sec:energies} Vertical and Relaxed Energies}

\begin{figure}[h]
	\includegraphics[width=0.95\linewidth]{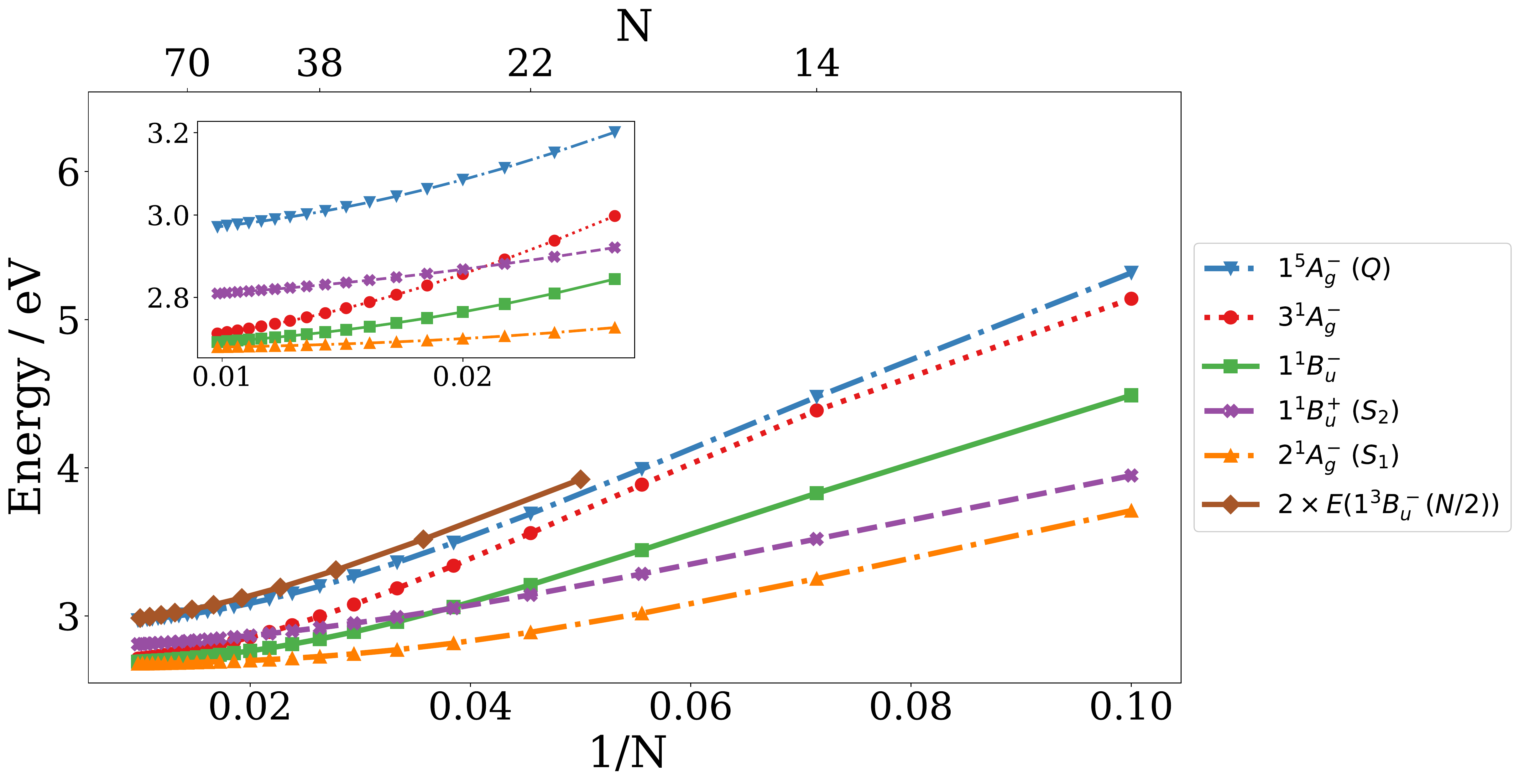}
	\caption{Vertical excitation energy of low-lying singlet and quintet states. Also shown is twice the vertical excitation energy of the triplet for chain lengths $N/2$. $N$ is the number of C-atoms. The insert shows the energies in the asymptotic limit.}
	\label{fig:vert}
\end{figure}

 \begin{figure}[h]
	\includegraphics[width=0.95\linewidth]{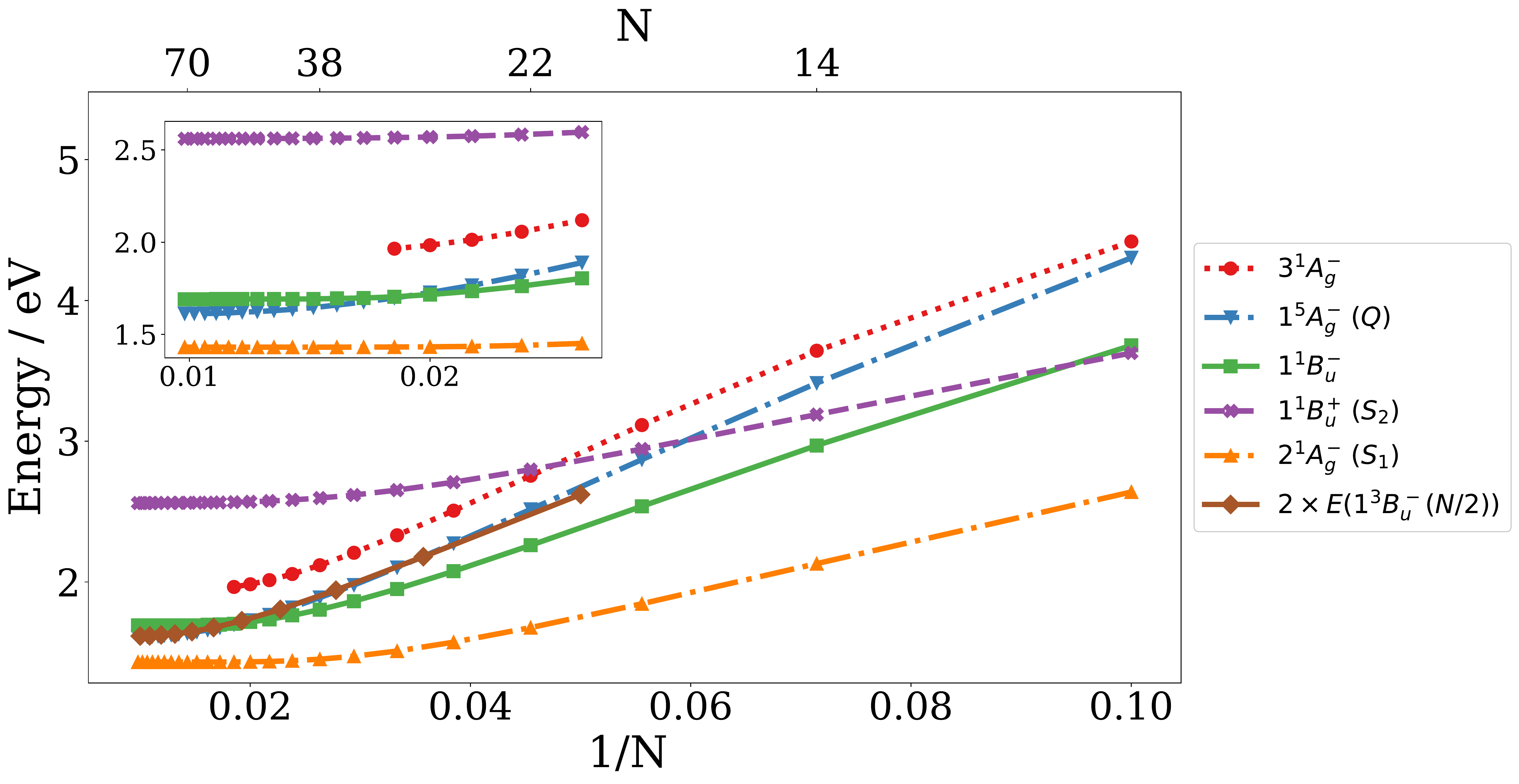}
		\caption{Relaxed excitation energy of low-lying singlet and quintet states. Also shown is twice the relaxed excitation energy of the triplet for chain lengths $N/2$. $N$ is the number of C-atoms. The insert shows the energies in the asymptotic limit.}
		\label{fig:relax}
\end{figure}

Using the Hellmann-Feynman iterative procedure, the relaxed geometries of the ground state, $1^1A_g^-$, for a range of chain lengths up to 102 carbon atoms (or sites) are calculated using the DMRG method \cite{Schollwock2005a,Barford2001,White1992}.

The vertical excitation energies for the lowest energy singlets are shown in  Fig.\ \ref{fig:vert}. For short chains we see the usual energetic ordering of $2^1A_g^- < 1^1B_u^+ <   1^1B_u^-  < 3^1A_g^-$  \cite{Tavan1987, Kurashige2004}. For chain lengths greater than 26 sites the vertical \bum \ energy becomes lower than the $1^1B_u^+$ energy, while at chain lengths greater than 46 sites the \agthree{} vertical energy falls below the \bu \ energy.

The $1^5A_g^-$ quintet state,  however, remains above the bright state at all chain lengths. Its energy converges to twice the triplet energy evaluated at half the chain length, implying that it corresponds to an unbound triplet-pair. This assumption will be confirmed by an analysis of the bond-dimerization and spin-spin correlation in the following sections. (This observation of the triplet-pair character of the quintet state is often invoked for simpler models, but remains valid with interacting electrons and electron-nuclear coupling \cite{Musser2019}.)

The inset of Fig.\ \ref{fig:vert} shows that the vertical energies of the $2^1A_g^-$,  $1^1B_u^-$ and $3^1A_g^-$ states converge to the same value in the asymptotic limit, being $\sim 0.3$ eV lower than the vertical quintet state. This result indicates that these vertical singlet states are different pseudo-momentum members of the same family of excitations, as described in more detail in Section \ref{sec:wf}. They have different energies because of their different center-of-mass kinetic energies, which vanishes in the long-chain limit.

Turning now to the relaxed energies,  as shown in Fig. \ref{fig:relax} we find that the relaxed $2^1A_g^-$ state is always lower in energy than the relaxed $1^1B_u^+$ state.  For chain lengths greater than 10 and 20 sites the \bum and \agthree states, respectively,  also fall below the $1^1B_u^+$ state. The quintet state undergoes a considerable  geometry relaxation compared to the \bu \ state and its energy falls below the bright state for $N>16$.

Comparing the vertical and relaxed energies, we find that between 10 and 26 sites the vertical \bum \ state lies above the vertical \bu \ state, but the relaxed \bum \ state  is below the relaxed \bu \ state; and  similarly for the \agthree{} \ state between 20 and 46 site chains.
Thus, our calculations suggest that there might exist internal conversion pathways to these states from the optically excited \bu \ state. In addition, if spin mixing is allowed relaxation pathways could also involve the $1^5A_g^-$ quintet state.
Based on the experimental observations that certain relaxation pathways becoming available only for mid-band or higher excitation \cite{Antognazza2010, Musser2019}, these pathways are likely to have a barrier.


The $2A_g^-$ state is found to be a bound state compared to two free (relaxed) triplets. Whilst endothermic singlet fission is possible \cite{Wilson2013,Swenberg1968}, this state is unlikely to be involved in singlet fission and instead offers an alternative relaxation pathway, as has been observed experimentally \cite{Musser2013, Antognazza2010}. Similarly, for realistic chain lengths the relaxed $1^1B_u^-$ energy lies below the relaxed energy of two free triplets, while the relaxed $3^1A_g^-$ energy lies above them for all chain lengths.
As for the vertical calculation, for the relaxed states we find that $E(1^5A_g^-) \approx 2 \times E(1^3B_u^-[N/2])$.


We note that the relaxation energies increase as  $3^1A_g^-  < 1^1B_u^- < 2^1A_g^- $. Consequently, unlike the vertical energies, the relaxed energies of the $2^1A_g^-$,  $1^1B_u^-$ and $3^1A_g^-$ states do not converge to the same value in the asymptotic limit, and indeed they saturate for $N \gtrsim 50$. This energy saturation occurs because of self-localization of the solitons, which is a consequence of treating the nuclei as classical variables and can be corrected by using a model of fully quantized nuclei \cite{Barford2002,Barford2013a}
In practice, however, in realistic systems disorder will also act to localise excited states \cite{Tozer2014}.


\section{\label{sec:bond_dime}Soliton Structures}

In the even $N$ polyene ground state, nuclei are dimerized along the chain, with a repeated short-long-short bond arrangement.
Electronically excited states lower their total energy by distorting from the ground state geometry. In some cases, the ground state bond alternations is reduced or reversed over sections of the chain. The change of dimerisation is characterised by domain walls called solitons \cite{Heeger1988,Barford2013a, Hayden1986, Barford2001,Roth2015, Su1995}.
 In neutral chains with an even number of sites, each soliton ($S$) is associated with an antisoliton ($\bar{S}$).

Solitons in linear conjugated systems are of two types: radical or ionic. For a radical soliton associated with covalent states the nuclei distortion is centered around a radical unpaired spin (or a spinon); the soliton has a net spin but is neutral.
For an ionic soliton, however, the distortion is associated with an unoccupied or doubly occupied site, so has  $S^z = 0$ but is charged \cite{Barford2013a,Roth2015}.

To investigate the solitonic structure of the excited singlet and quintet states, we calculate the staggered, normalised bond dimerization, $\delta_n$, of their relaxed geometries, defined as
\begin{equation}
    \delta_n = (-1)^n \frac{(t_n - \bar{t})}{\bar{t}},
\end{equation}
with $t_n = t_0 + \alpha (u_{n+1}- u_n)$  and $\bar{t}$ being the average of $t_n$. For the ground state $\delta_n \approx 0.85$, across the chain with the bond dimerization being slightly larger at the ends of the chain.

The lowest lying triplet state, \triplet{}, is a two-soliton state (i.e., $S\bar{S}$) with each soliton being associated with a radical spin, residing towards the ends of the chain. On the other hand, the $2^1A_g^-$, \bum \ and $1^5A_g^-$ states  are four-soliton states.

\begin{figure}[h]
    \centering
    \begin{subfigure}[t]{\linewidth}
        \centering
        \xincludegraphics[width=.8\linewidth, label=(a)]{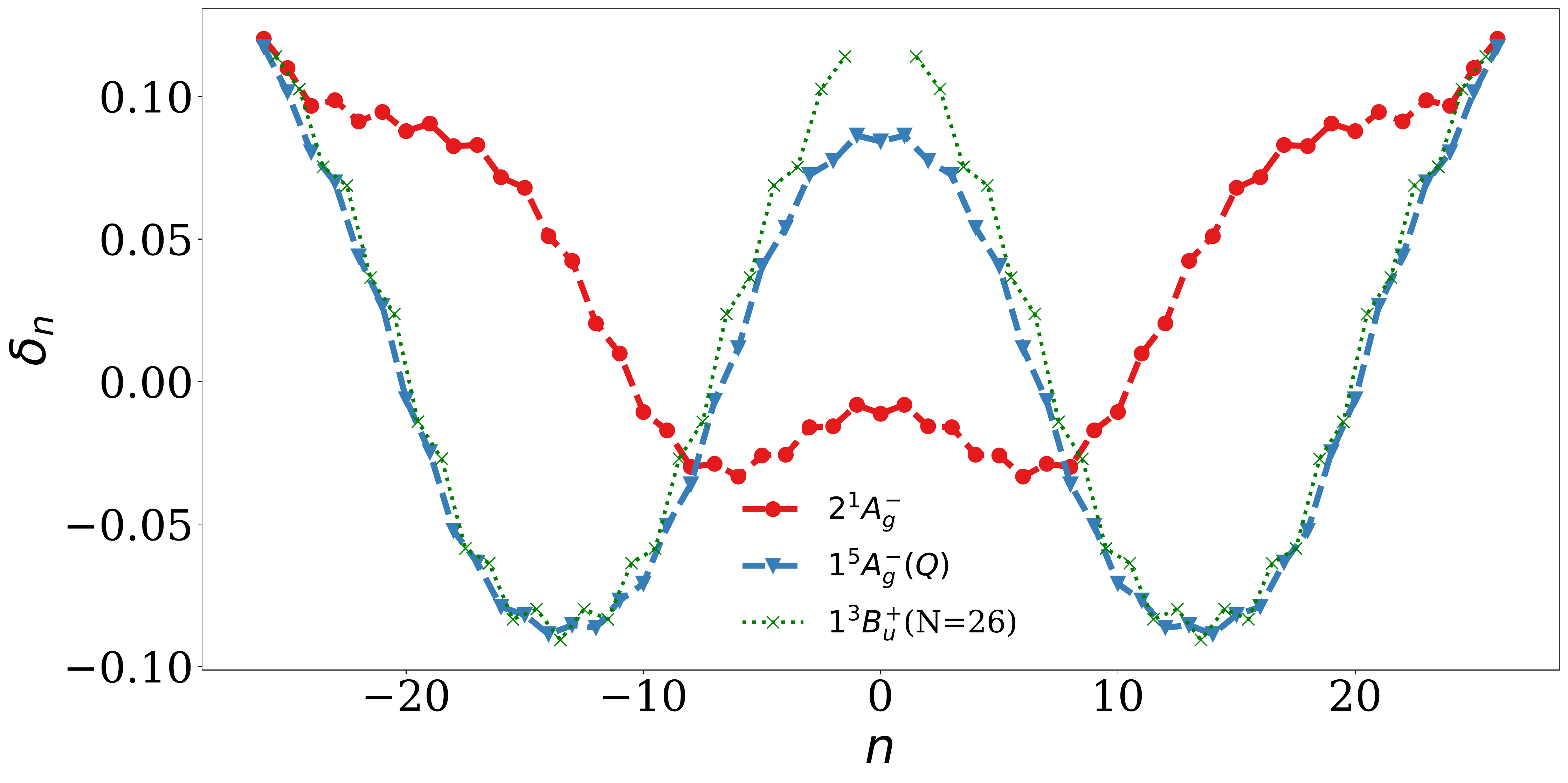}
		\label{fig:bondorder3Ag2Ag}
    \end{subfigure}
    \begin{subfigure}[b]{\linewidth}
        \centering
        \xincludegraphics[width=.8\linewidth, label=(b)]{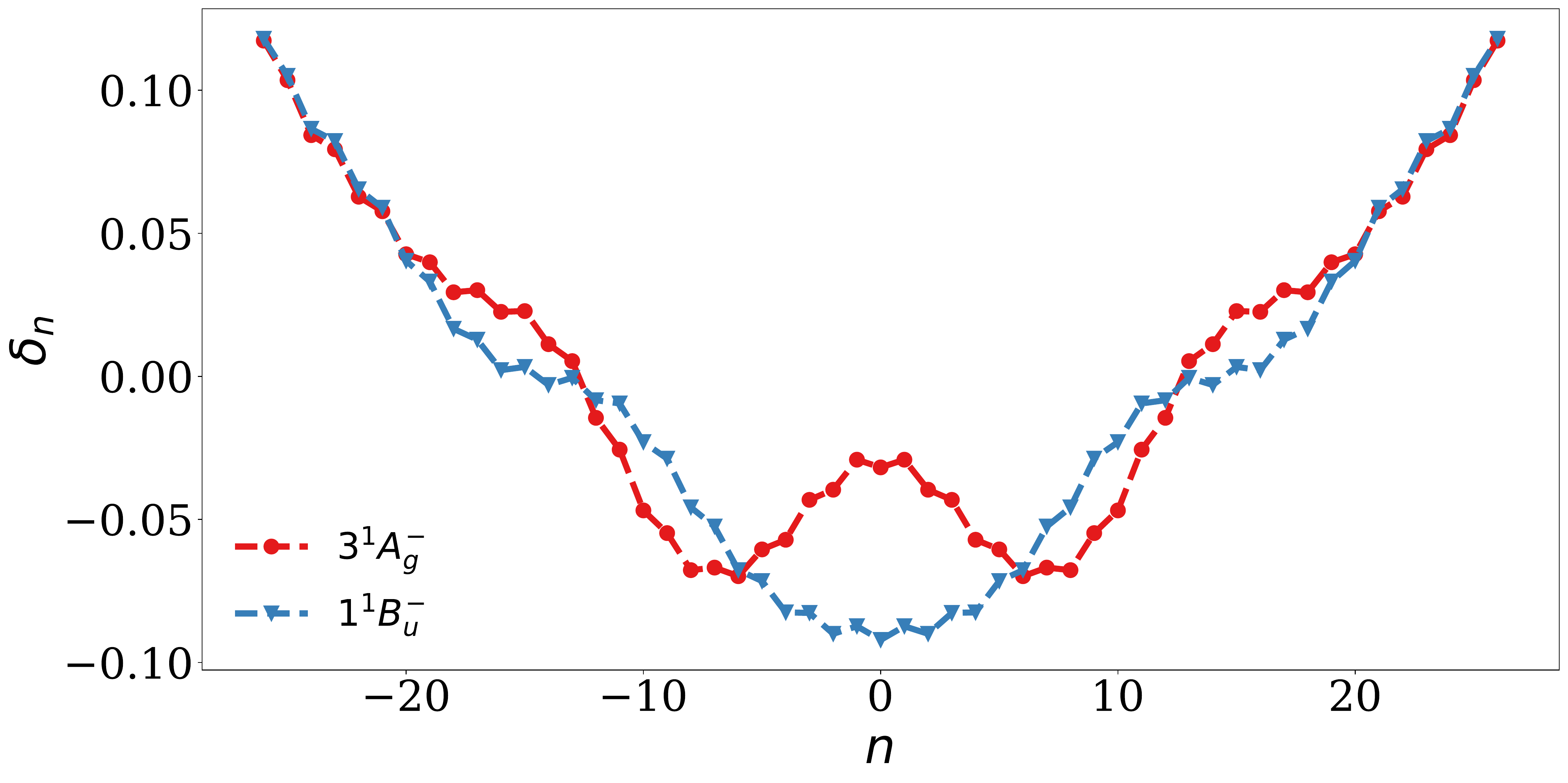}
    	\label{fig:bond_order_1BuM}
    \end{subfigure}
    \caption{The staggered bond dimerization, $\delta_n$ for a chain of 54 sites, of the (a) the \agtwo \ , $1^5A_g^-$\ and \triplet \ (for $N=26$) states and (b) the \bum \ and \agthree  states.}
        \label{fig:bond_order_4}
\end{figure}

Fig.\ \ref{fig:bond_order_4}(a) presents the staggered bond dimerization for the $2^1A_g^-$ and $1^5A_g^-$ states, implying that the soliton arrangement is  $S\bar{S} S\bar{S}$.
The $1^5A_g^-$ \ bond dimerization strongly resembles that of two triplets residing on either half of the chain, suggesting that the $1^5A_g^-$ \ state consists of two spatially separated triplets.
The bond dimerization of the \agtwo \ state is well-known \cite{Hu2015,Hayden1986,Tavan1987,Heeger1988,Su1995} \cite{Barford2001}: the solitons are more bound, indicating that the \agtwo{} state is a bound triplet-pair. These observations are quantified by fitting the bond dimerization of the \agtwo \ and $1^5A_g^-$ \ states by\cite{Su1995}
\begin{widetext}
\begin{equation}
\label{eqn:four_soliton}
\begin{split}
 \delta_n = \delta_0 \Big[ 1 + \tanh \left( \frac{2 n_0 a}{\xi} \right) \Big\lbrace \tanh \left( \frac{2 (n-n_d-n_0) a}{\xi} \right)  - \tanh \left( \frac{2 (n-n_d+n_0) a}{\xi} \right) \\
 + \tanh \left( \frac{2 (n+n_d-n_0) a}{\xi} \right)  - \tanh \left( \frac{2 (n+n_d+n_0) a}{\xi} \right)\Big\rbrace \Big],
 \end{split}
\end{equation}
\end{widetext}
where $\xi$ is the domain wall width, $2n_0$ is the separation of the soliton and antisoliton within a $S\bar{S}$ pair on either side of the chain, while $2n_d$ is  the separation of the pairs.

The  bond dimerization of the \bum{} state, shown in Fig.\ \ref{fig:bond_order_4}(b), can be explained by the soliton arrangement of $SS\bar{S}\bar{S}$. Its bond dimerization fits the equation
\begin{widetext}
\begin{equation}
\label{eqn:four_soliton_alt}
\begin{split}
 \delta_n = \delta_0 \Big[ 1 + \frac{1}{2} \tanh \left( \frac{2 n_0 a}{\xi} \right)\Big\lbrace \tanh \left( \frac{2 (n-n_d-n_0) a}{\xi} \right)  + \tanh \left( \frac{2 (n-n_d+n_0) a}{\xi} \right) \\
 - \tanh \left( \frac{2 (n+n_d-n_0) a}{\xi} \right)  - \tanh \left( \frac{2 (n+n_d+n_0) a}{\xi} \right)\Big\rbrace \Big],
 \end{split}
\end{equation}
\end{widetext}
where $2n_0$ is the separation of the soliton and soliton within a $SS$ pair (and likewise of the antisoliton and antisoliton within a $\bar{S}\bar{S}$ pair), while $2n_d$ is  the separation of these pairs.
The \agthree state{} bond dimerization can be explained by the six-soliton arrangement of $SS\bar{S}S\bar{S}\bar{S}$. As we will see in Section \ref{sec:triplet_triplet}, there are many different triplet-triplet contributions to the \agthree{} state, which when summed give rise to a more complicated bond dimerization.

\begin{figure}[h]
\centering
    \begin{subfigure}[b]{\linewidth}
        \xincludegraphics[width=0.8\textwidth,label=(b)]{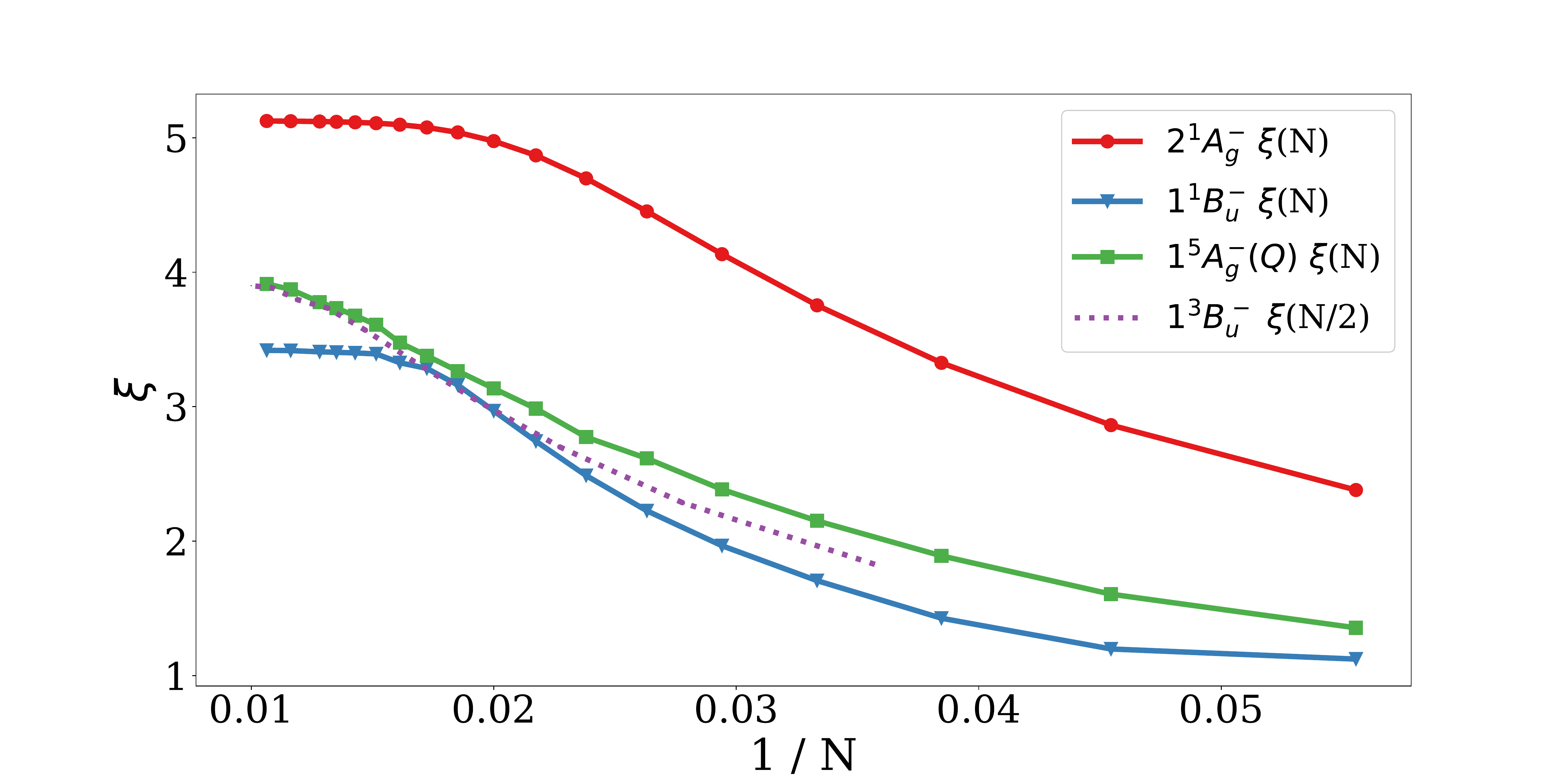}
        \label{fig:fitted_parameters_2}
    \end{subfigure}
        \begin{subfigure}[b]{\linewidth}
        \xincludegraphics[width=0.8\textwidth,label=(a)]{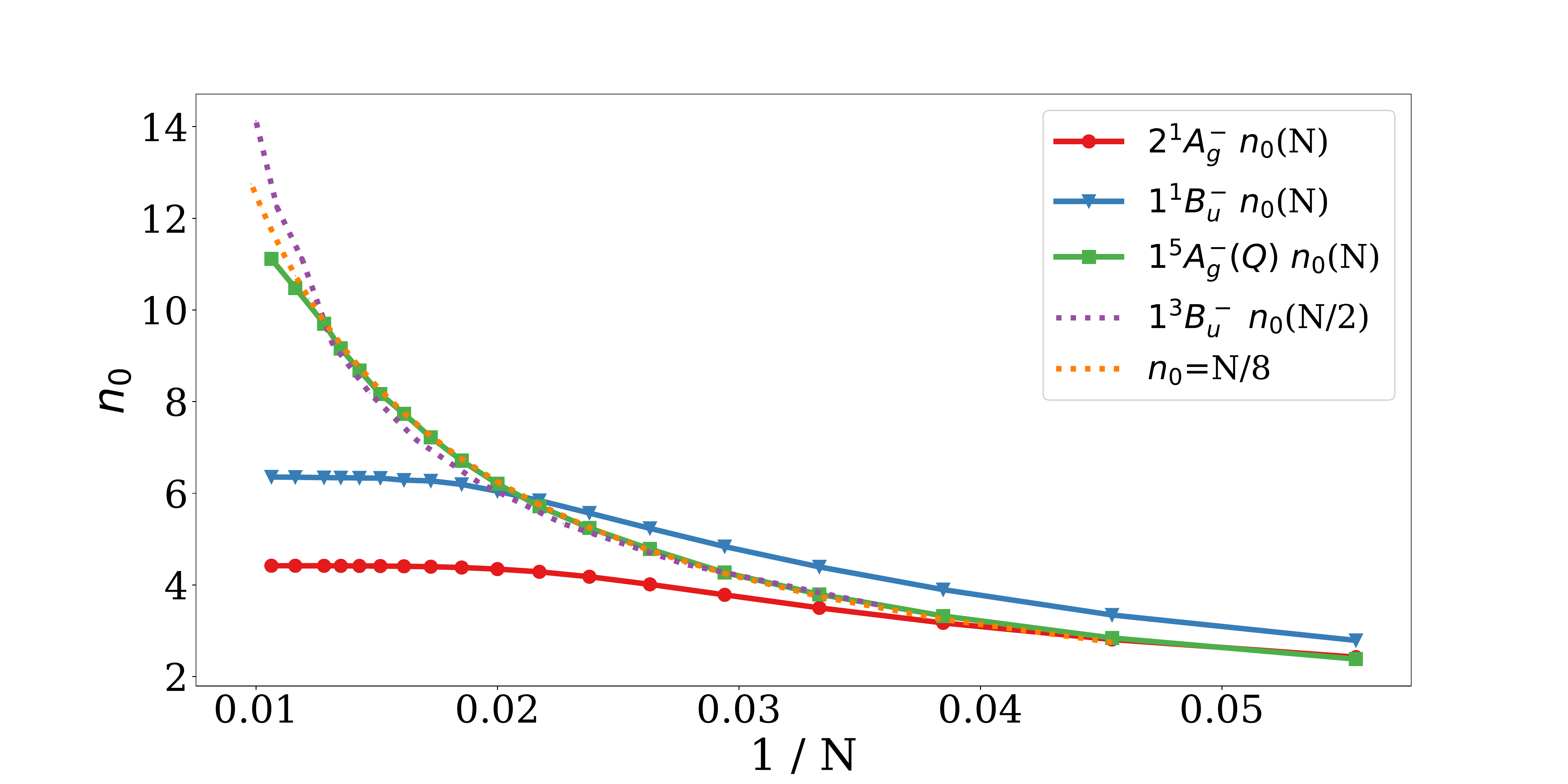}
        \label{fig:fitted_parameters_1}
    \end{subfigure}
    \begin{subfigure}[b]{\linewidth}
        \xincludegraphics[width=0.8\textwidth,label=(c)]{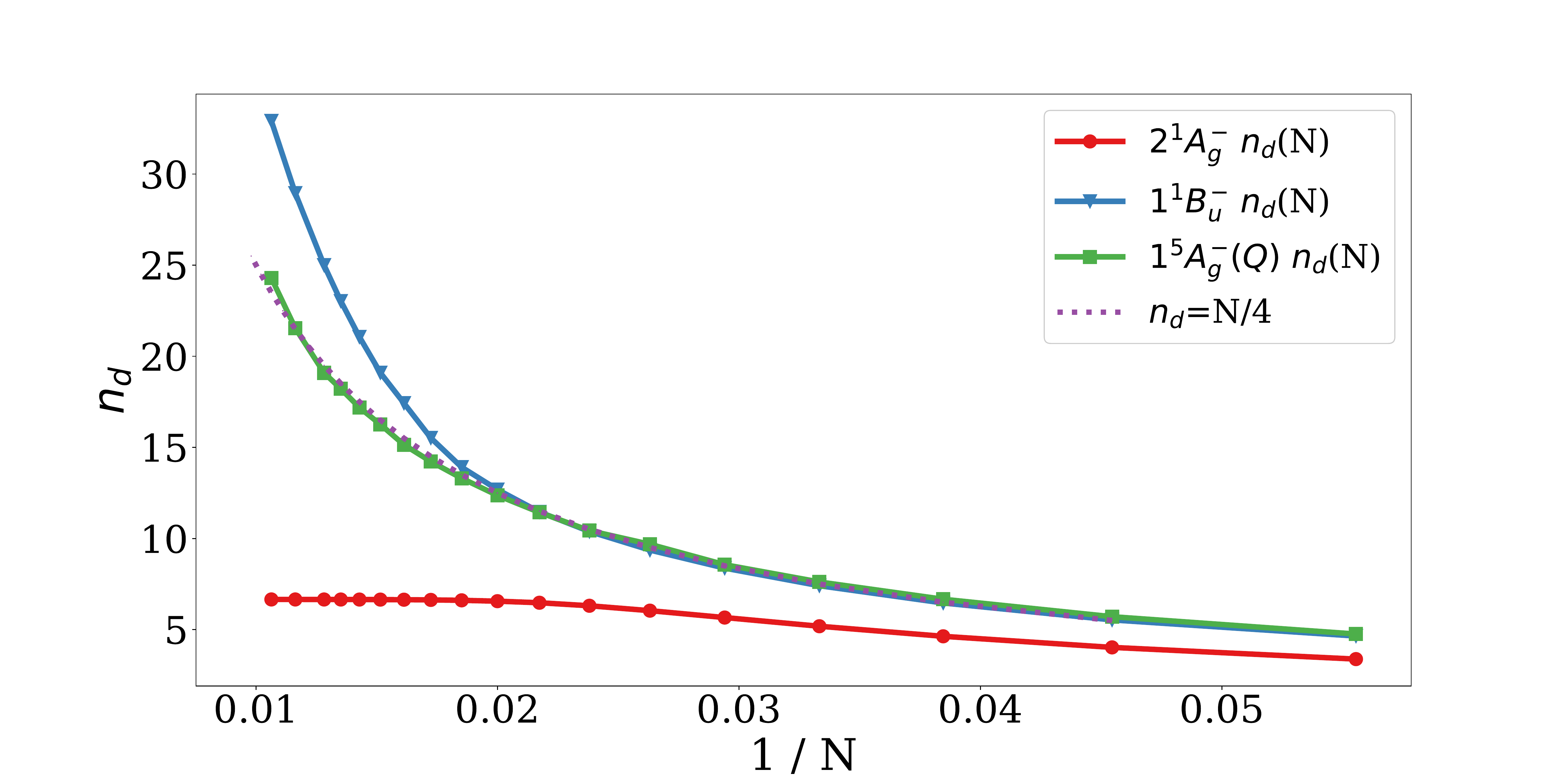}
        \label{fig:fitted_parameters_3}
        \end{subfigure}
         \caption{The fitted parameters, (a) $\xi_i$, (b) $n_0$, and (c)  $n_d$ plotted against inverse chain length for the four-soliton states \agtwo, \Q{} \  and \bum. For comparison, in (a) the \triplet \ $n_0$ for length $N/2$, in (b) the \triplet \ $\xi$ for length $N/2$ is plotted and in (c) the curve $n_d= N/4$ is plotted.}
\label{fig:fitted_parameters}
\end{figure}

The fitted parameters, $n_0$, $n_d$ and $\xi$, for the three four-soliton states are plotted in Fig.\ \ref{fig:fitted_parameters} against inverse chain length. In Fig.\ \ref{fig:fitted_parameters}(a) we see that the coherence length, $\xi$, converges with chain length for all states.

The rapid convergence  of $n_0$ with chain length for the \agtwo \ state, shown in Fig. \ref{fig:fitted_parameters}(b), implies that the solitons within a $S\bar{S}$ pair are more strongly bound compared to other states. For the $1^5A_g^- $ state,  the change in $n_0$ with chain length, $N$, resembles that of the \triplet \ state for a chain of half-length, $N/2$, indicating that the $1^5A_g^- $  state has significant $T_1-T_1$ character with two triplet-like excitations occupying either side of the chain. $n_0$ converges as a function of $N$ for the \bum \ state, implying that the solitons (antisolitons) within a $SS$ ($\bar{S}\bar{S}$) pair are bound.

The distances between soliton pairs, $n_d$, are shown in Fig.\ \ref{fig:fitted_parameters}(c). Again, for the \agtwo \ state there is rapid convergence in the separation of these pairs. In contrast, both the $1^5A_g^-$ and \bum \ states do not show convergence, with the pair separation increasing as the chain length increases.
The $S\bar{S}$ pair distance for the $1^5A_g^-$ state follows $n_d \approx N/4$, again indicating that the pairs are unbound.

\section{\label{sec:spin}Spin-spin Correlation}

In addition to the bond dimerization, further insight into the radical (spinon) character of the triplet-pair states is obtained via the spin-spin correlation function, defined as
\begin{equation}
S_{nm} = \Braket{S_n^z S_m^z}.
\end{equation}
A positive/negative spin-spin correlation value indicates a ferromagnetic/antiferromagnetic alignment between a pair of spins.

\begin{figure*}[h]
    \centering
    \begin{subfigure}[h]{0.45\linewidth}
        \includegraphics[width=\textwidth]{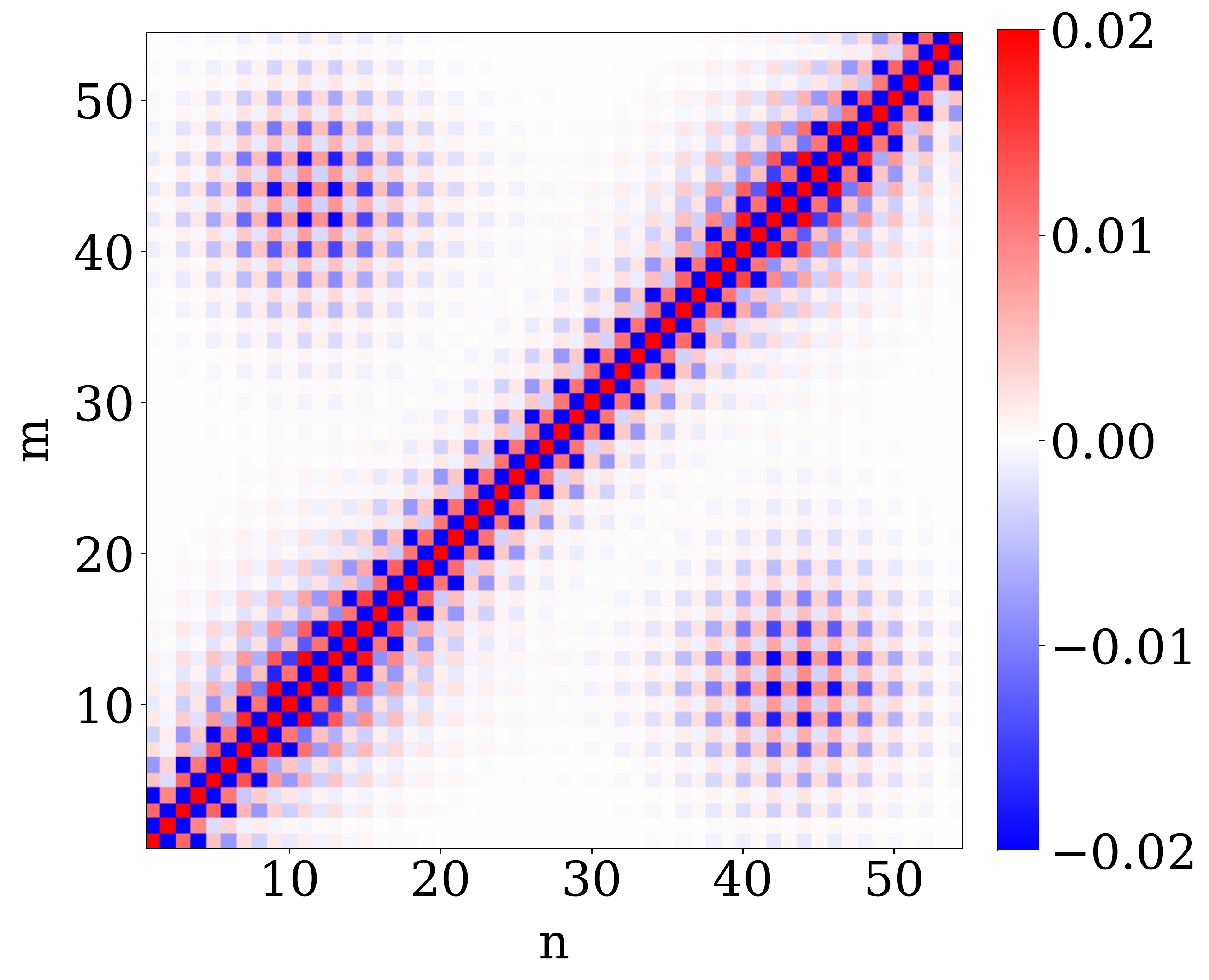}
        \caption{\triplet}
        \label{fig:spin_spin_T}
    \end{subfigure}
    \centering
    \begin{subfigure}[h]{0.45\linewidth}
        \includegraphics[width=\textwidth]{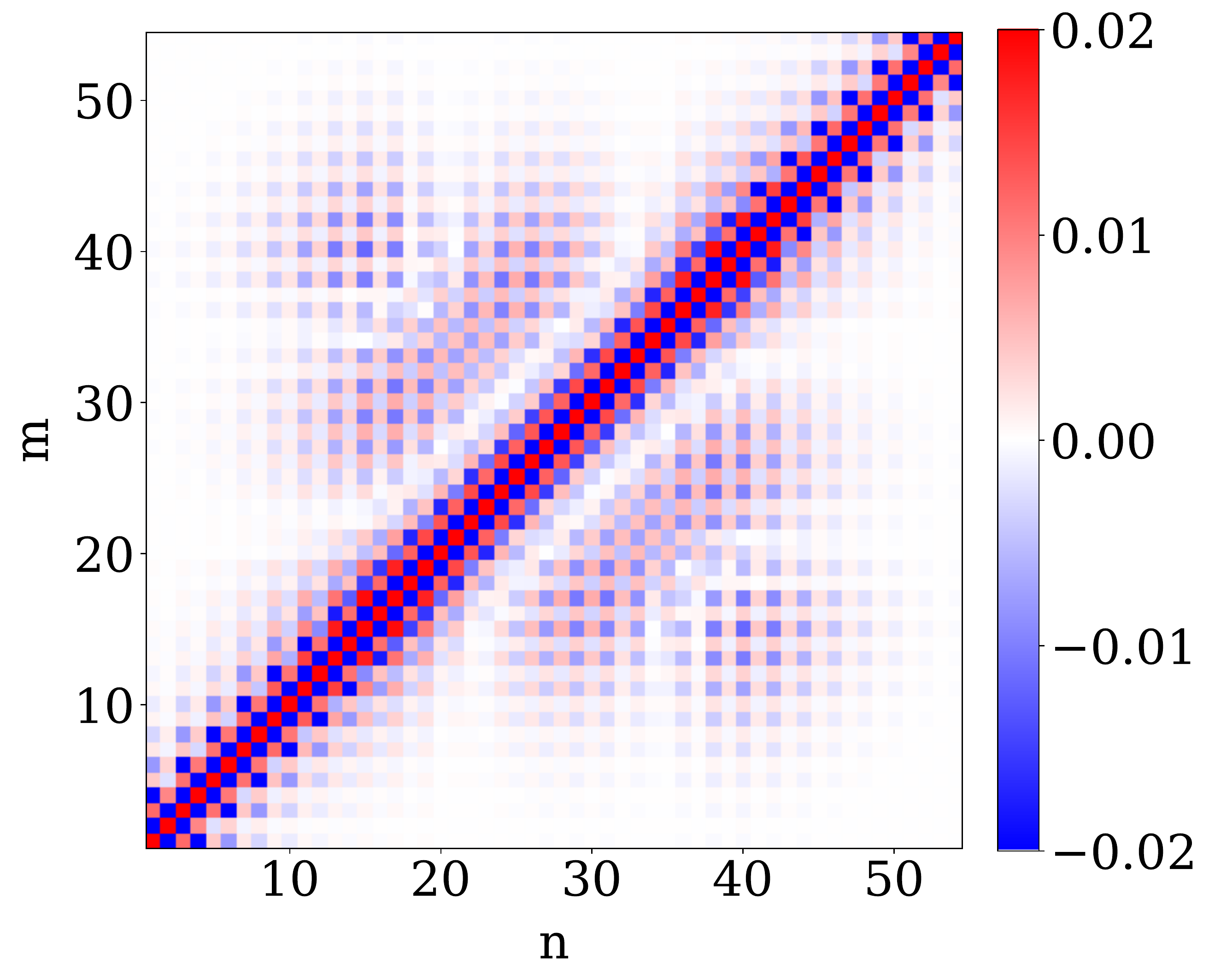}
        \caption{\agtwo}
        \label{fig:spin_spin_2Ag}
    \end{subfigure}
    \begin{subfigure}[h]{0.45\linewidth}
        \includegraphics[width=\textwidth]{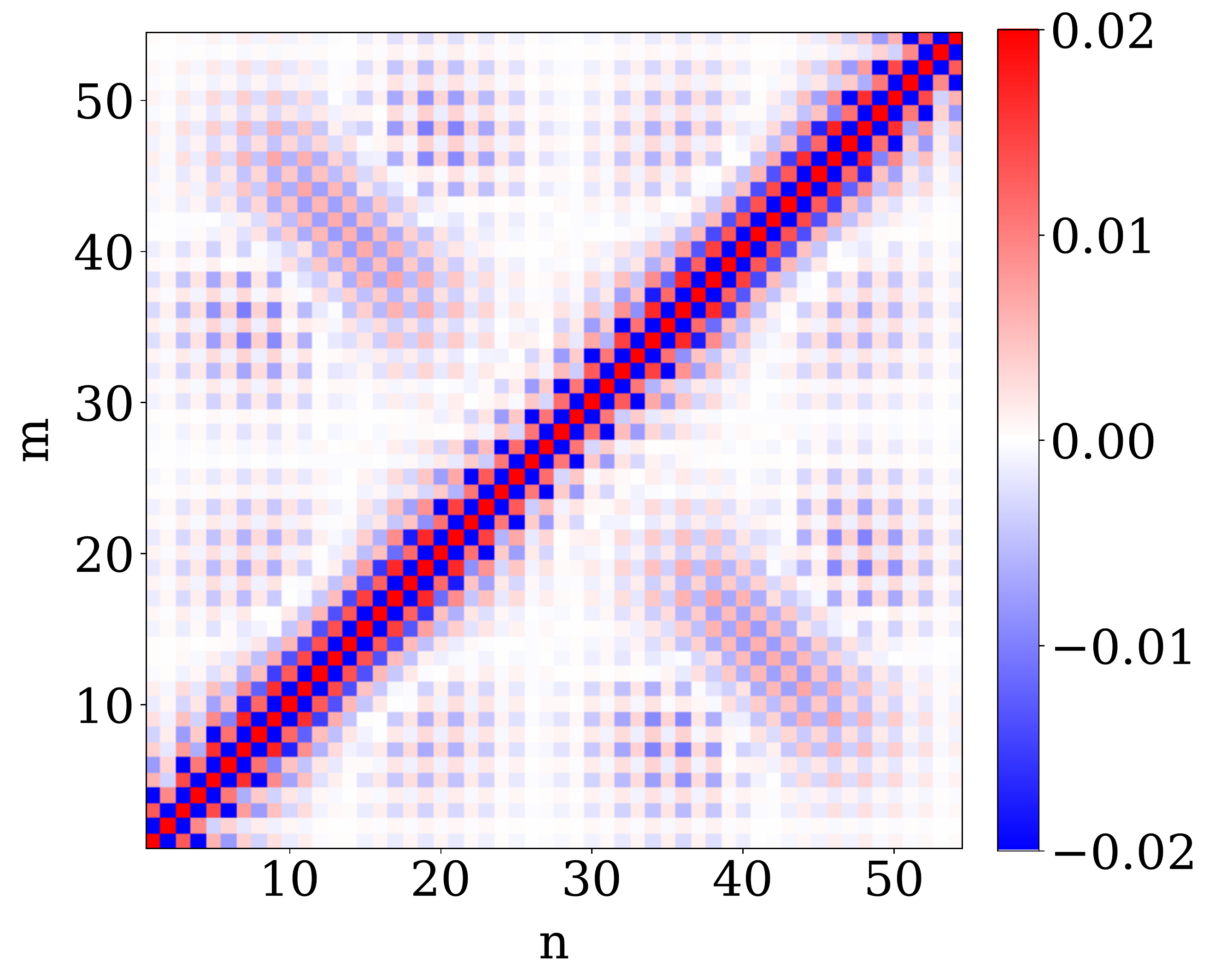}
        \caption{\bum}
        \label{fig:spin_spin_1BuM}
        \end{subfigure}
           \begin{subfigure}[h]{0.45\linewidth}
        \includegraphics[width=\textwidth]{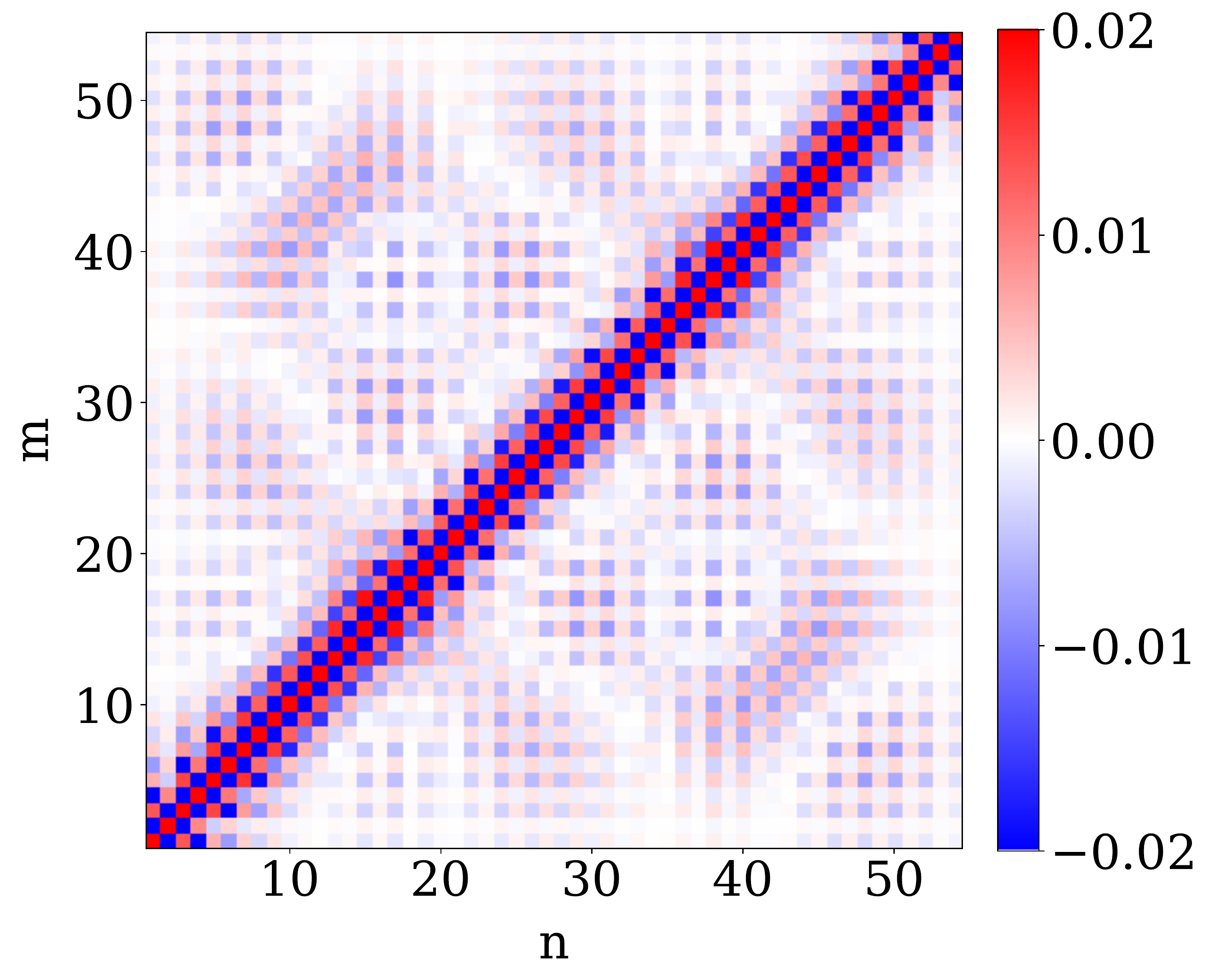}
        \caption{\agthree}
        \label{fig:spin_3Ag}
        \end{subfigure}
            \begin{subfigure}[h]{0.45\linewidth}
        \includegraphics[width=\textwidth]{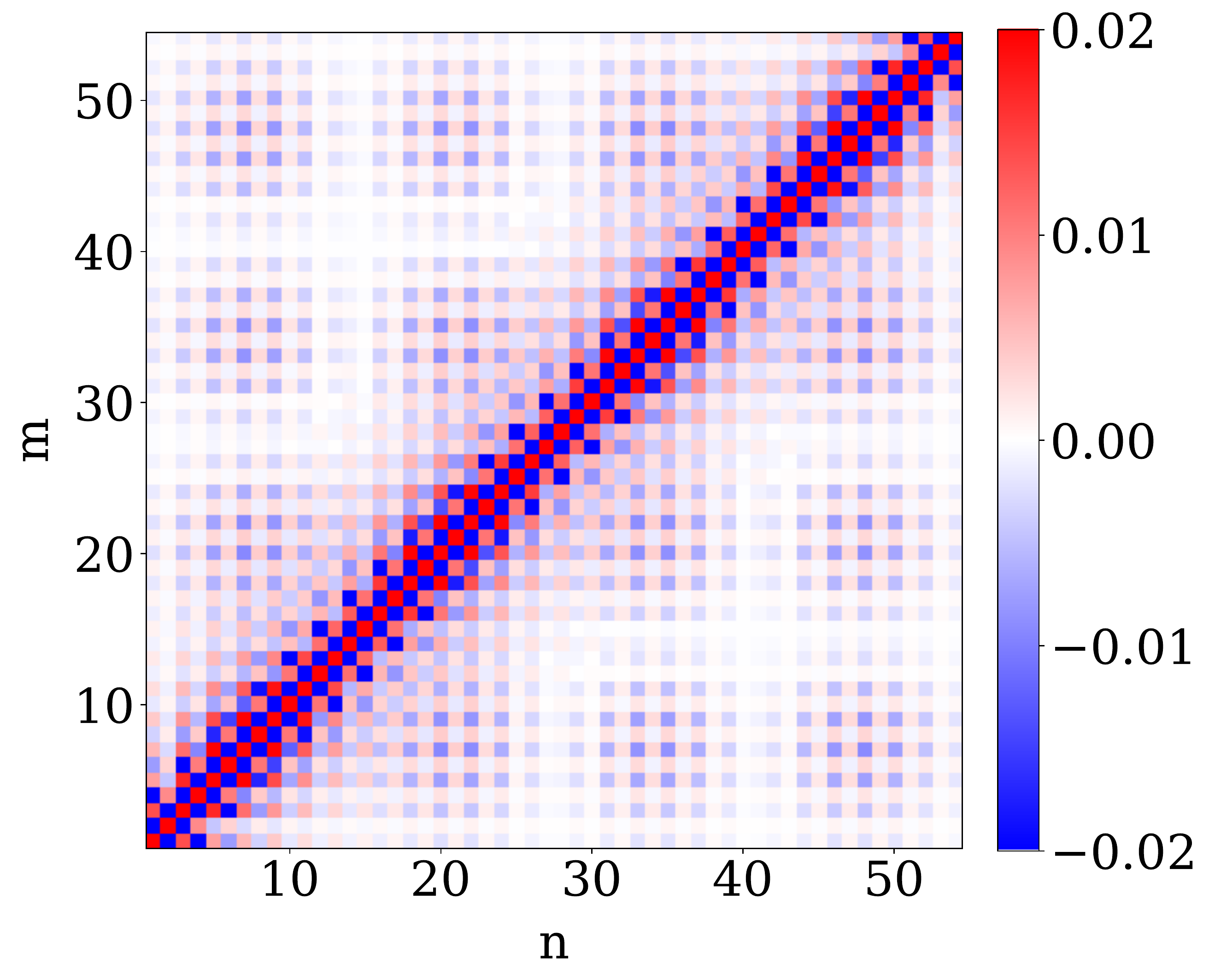}
        \caption{$1^5A_g^-$}
        \label{fig:spin_spin_Q}
    \end{subfigure}
    \begin{subfigure}[h]{0.45\linewidth}
        \includegraphics[width=\textwidth]{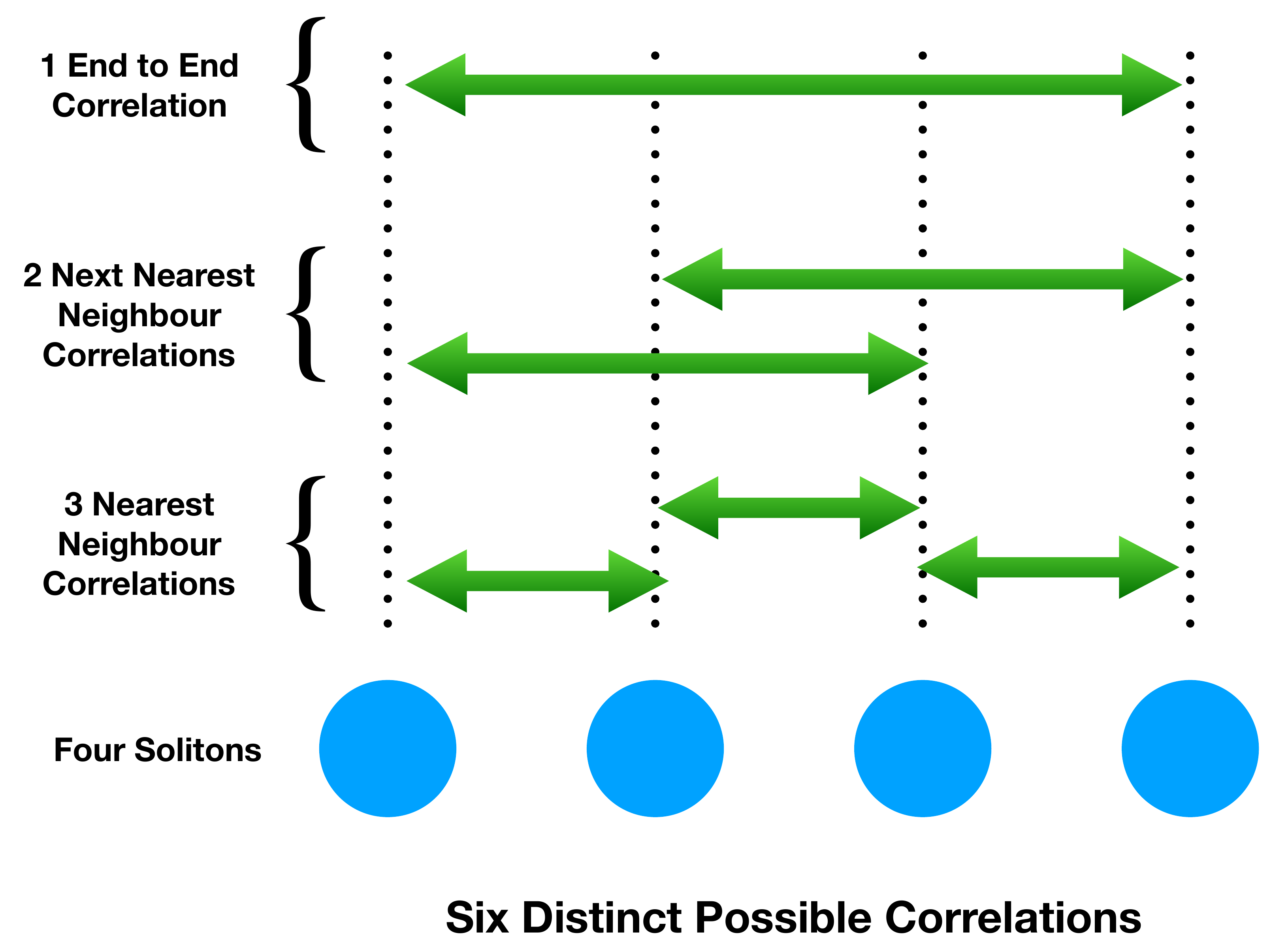}
        \caption{}
        \label{fig:spin_spin_corr}
        \end{subfigure}
        \caption{The  spin-spin correlations for the relaxed (a) \triplet, (b) \agtwo{},  (c) \bum{},  (d) \agthree{}, and (e) \Q{} states. (f) shows a  schematic of the correlations from the four spin-radicals (solitons) of the \Q{} \ state. (Note that  the values of $S_{nm}$ for $n = m$ to $n=m\pm4$ are larger than scale provided. The scale presented is used to emphasize the long range correlations.)}
\label{fig:spin_spin}
\end{figure*}

The  spin-spin correlations for the relaxed \triplet{} state are shown in Fig.\ \ref{fig:spin_spin_T}. We see that  the radical soliton/antisoliton of the triplet state localize towards the end of the chain and there is a long range spin-spin correlation between them. The solitons are delocalized over a small region well described by the coherence length, $\xi$.

Fig.\ \ref{fig:spin_spin_Q} shows the  spin-spin correlation for the relaxed \Q{} state. We see three correlations between neighbouring solitons and antisolitons,  and three further long range correlations. This correlation pattern is consistent with two unbound triplets, i.e.,  four solitons positioned along the chain as predicted from Eq.\ \ref{eqn:four_soliton} and presented in Fig.\ \ref{fig:bond_order_4}a. A schematic of the soliton interactions that lead to these six correlations is shown in Fig.\ \ref{fig:spin_spin_corr}.

Fig.\ \ref{fig:spin_spin_2Ag} shows $S_{nm}$ for the \agtwo{} state.
It is difficult to discern correlations between individual solitons, because the correlations overlap each other. However, along the anti-diagonal, $m = (N-n)$,  long range correlations  between sites $\approx 10 $ and $ \approx 40$ can be seen. Overall, the spin-spin correlations of the \agtwo{} state further confirm its bound triplet-pair character. The triplets are bound in the middle of the chain,  and individual solitons contributing to the triplets cannot be identified.

$S_{nm}$ for the relaxed \bum \ state shows correlations similar to  \agthree{}, but with much more delocalized correlations along the antidiagonal.

\section{Excited State Wavefunctions}\label{sec:wf}

The low-lying singlet dark states, i.e., $2^1A_g^-$, $1^1B_u^-$ and $3^1A_g^-$, have negative particle-hole symmetry and are sometimes characterised as being `covalent' or of predominately spin-density-wave (SDW) character. In contrast, the optically allowed $1^1B_u^+$ state has positive particle-hole symmetry and is characterised as being `ionic' or of electron-hole character. (In this paper we adopt the chemists' definition of particle-hole symmetry, which is opposite to the physicists' notation (see ref. 17).)

In practice, however, the multi-excitonic $2^1A_g^-$, $1^1B_u^-$ and $3^1A_g^-$ states have both covalent and ionic character. In addition, as we saw in Section \ref{sec:energies}, the vertical energies of these states converge to the same value as $N \rightarrow \infty$ suggesting that they are related. In this section we describe the multi-excitonic character of these states and explain how they are members of the same family of excitations.

We first discuss the triplet-pair components of these states before describing their excitonic wavefunctions.

\subsection{\label{sec:triplet_triplet} Triplet-Triplet Overlap}

By comparing the excitation energy of the low-energy singlets of polyenes with the excitation energy of the individual triplets, it has been suggested that the triplet-triplet combinations that contribute to each state are:  \cite{Taffet2019, Tavan1987}
\begin{equation}
	\begin{split}
	2^1A_g^- \equiv T_1 \otimes T_1 \\
	1^1B_u^- \equiv T_1 \otimes T_2 \\
	3^1A_g^- \equiv T_2 \otimes T_2 \\
	\end{split}
\end{equation}
where $T_1 \equiv 1^3B_u^-$   and $T_2 \equiv 1^3A_g^-$.

To quantify the triplet-triplet character of the \agtwo, \agthree \ and \bum{} singlet states we compute their overlap with triplet-pair direct product wavefunctions. We calculate the vertical excited state wavefunctions for a chain of 12 sites.
Then, by calculating the triplet states for a chain of 6 sites and taking the direct product of different pairs of triplets, we can generate states that have triplets on either half of the chain.
The square overlap of the triplet-triplet wavefunctions with the \agtwo, \agthree \ and \bum \ states are presented in Tab.\ \ref{tab:2Ag_overlap}.

The \agtwo{} state, whilst primarily consisting of $T_1 \otimes T_1$ components, also contains some $T_1 \otimes T_2$ character.  The \bum \ state consists exclusively of $T_1 \otimes T_2$, as only these combinations are symmetry allowed.  The \agthree{} state, rather than primarily having $T_2 \otimes T_2$ character, has both $T_1 \otimes  T_1$ and symmetry allowed combinations of $T_1$ and $T_2$ components in its wavefunction. Indeed,  the sum of the $T_1 \otimes  T_1$ and $T_1 \otimes  T_2$  components has larger amplitude than  the $T_2 \otimes  T_2$ character. Since the \agthree{} state has character from each of the triplet-triplet combinations, the sum of these contributions lead to the complicated staggered bond dimerization and spin-spin correlation, discussed in Section \ref{sec:bond_dime} and Section \ref{sec:spin}.


\renewcommand{\arraystretch}{1.4}
\begin{table}
\centering
\begin{tabular}{ C{1.2cm}  C{1.2cm}  C{2cm}   C{2cm}   C{2cm}  }
\toprule
  &  & \multicolumn{3}{c}{ $|\Braket{ T_l \otimes T_r | \Psi} |_{v} ^2$}\\
$T_l$ & $T_r$ & $ 2^1A_g^-$ & $ 3^1A_g^- $ &  $1^1B_u^-$ \\
\midrule
 $T_1^{0}$ & $T_1^{0}$ &  0.134 & 0.020  & -\\
 $T_1^{+1}$ & $T_1^{-1}$ &  0.134 & 0.020 & -\\
 $T_1^{-1}$ & $T_1^{+1}$ &  0.134 & 0.020 & -\\
 $T_1^{0}$ & $T_2^{0}$ &  0.010 & 0.012 & 0.022 \\
 $T_2^{0}$ & $T_1^{0}$ &  0.010  & 0.012 & 0.022\\
 $T_1^{+1}$ & $T_2^{-1}$ &  0.010 & 0.012 & 0.022\\
 $T_2^{-1}$ & $T_1^{+1}$ &  0.010 & 0.012 & 0.022 \\
 $T_2^{+1}$ & $T_1^{-1}$ &  0.010 & 0.012 & 0.022 \\
 $T_1^{-1}$ & $T_2^{+1}$ &  0.010 & 0.012 & 0.022\\
 $T_2^{0}$ & $T_2^{0}$ &  - & 0.015 & - \\
 $T_2^{+1}$ & $T_2^{-1}$ & - & 0.015 & - \\
 $T_2^{-1}$ & $T_2^{+1}$ &  - & 0.015 & -\\
 \midrule
 \multicolumn{2}{c}{Total} & 0.462 & 0.177  &  0.132\\
 \bottomrule
\end{tabular}
\caption{Table of the square of overlaps, $|\Braket{ T_l \otimes T_r | \Psi} | ^2$ for the vertical  \agtwo{}, \bum{} and \agthree{} states. $T_r$ and $T_l$ are calculated for a 6 site chain (the super script indicates the $S_z$ eigenvalue of the state, $T_1$ and $T_2$ corresponds to the $1^3B_u^-$ and $1^3A_g^-$ triplets, respectively), while the state $\Psi$ is calculated for an 12 site chain.}
\label{tab:2Ag_overlap}
\end{table}

\renewcommand{\arraystretch}{1}

\subsection{Triplet-pair Wavefunctions}\label{sec:bimagnon}

In the previous section we saw that higher-energy covalent states are composed of linear combinations of higher-energy triplet states. In this section we quantify how to construct bound triplet-pair states from free triplet-pair states. To do this it is convenient to assume translationally invariant systems. We also assume a dimerized antiferromagnetic groundstate, from which bound triplet-pair excitations are predicted. \cite{Harris1973,Uhrig1996}

Suppose that $a_{k_1}^{\dagger}$ and $a_{k_2}^{\dagger}$  create  triplet excitations (or more precisely, bound spinon-antispinon pairs\cite{Uhrig1996}) with wavevectors $k_1$ and $k_2$. Then a free triplet-pair  excitation is
\begin{equation}
\ket{k_1,k_2} = \ket{K-k’/2,K+k’/2} = a_{k_1}^{\dagger} a_{k_2}^{\dagger} \ket{\mathrm{GS}},
\end{equation}
where $2K = (k_1+k_2)$ is a the center-of-mass wavevector, $k’=(k_1-k_2)$ is the relative wavevector, and $\ket{\mathrm{GS}}$ represents the dimerized antiferromagnetic groundstate.

A bound triplet-pair excitation is a linear combination of the kets $\{\ket{k_1,k_2} \}$, namely
\begin{equation}
\ket{\Phi_n(K)} =  \sum_{k’,K’} \Phi_n(k’,K’) \ket{K-k’/2,K+k’/2},
 \end{equation}
where $\Phi_n(k’,K’)$ is the triplet-pair wavefunction in $k$-space and $n$ is the principal quantum number. $K$ is a good quantum number for the bound state (although $k'$ is not), hence
\begin{equation}
\Phi_n(k’,K’) =\psi_n(k’) \delta(K’-K)
 \end{equation}
and thus
\begin{equation}\label{eq:54}
\ket{\Phi_n(K)} =  \sum_{k’} \psi_n(k’) \ket{K-k’/2,K+k’/2}.
 \end{equation}
Fourier transforming $\Phi_n(k’,K’)$ gives the real-space triplet-pair wavefunction:
\begin{equation}
\tilde{\Phi}_{n,K}(r,R) = \tilde{\psi}_n(r)\tilde{\Psi}_K(R)
\end{equation}
where the center-of-mass wavefunction is the Bloch state
\begin{equation}
\tilde{\Psi}_K(R) = \frac{1}{\sqrt{N}}\exp(iKR).
\end{equation}
$R$ is the center-of-mass coordinate, and $\tilde{\psi}_n(r)$ is the relative wavefunction with $r$ being the T-T separation.

Equation (\ref{eq:54}) indicates that the bound triplet-pair state  is constructed from a linear combination of free triplet-pair states with different $k_1$ and $k_2$, subject to  a definite center-of-mass wavevector (and momentum). These states form a band, whose band width is determined by their center-of-mass kinetic energy. A bound triplet-pair is unstable to dissociation (or fission) if its  kinetic energy is greater than the triplet-pair binding energy.

\subsection{Exciton Wavefunctions}\label{sec:excitons}

We now describe the exciton wavefunctions of the low-energy states of linear polyenes, using a real-space representation.

The excitation of an electron from the valence-band to the conduction-band in semiconductors creates a positively charged hole in the valence-band. In conjugated polymers, the electrostatic interaction between the two create a bound electron-hole
pair, termed an exciton. Assuming the `weak-coupling' limit, excitons in conjugated polymers are described by an effective H-atom model \cite{Barford2002b,Barford2013a,Barford2013b} or by a mapping from a single-CI calculation \cite{Barford2008}. An excitation from the valence-band to the conduction-band can thus be characterized by an effective particle model \cite{Barford2013a}. In the real-space picture, an exciton is described by the center-of-mass coordinate of the exciton, $R$ and the relative coordinate, $r$\cite{Barford2013a}.  $r$ is a measure of the size of the exciton. The electron-hole coordinate, $r$, is associated with the principal quantum number, $n$ \cite{Barford2002b,Barford2013a,Barford2013b} , while the center-of-mass coordinate is associated with the center-of-mass quantum number, $j$.

We denote an exciton basis state by $\ket{R+r/2,R-r/2}$. The exciton creation operator $S_{rR}^\dagger$ creates a hole in the
valence-band orbitals at $(R-r/2)$ and an electron in the conduction-band orbitals at $(R+r/2)$, i.e.,
\begin{align}
	\ket{R+r/2,R-r/2}
		&=
			S_{rR}^\dagger \ket{\mathrm{GS}},
\end{align}
where $\ket{\mathrm{GS}}$ is the ground state in this basis.

To investigate the electron-hole nature of the excited states, we express an excited state $\ket{\Phi}$ as a linear
combination of the real-space exciton basis $\{\ket{R+r/2,R-r/2\}}$:
	\begin{align}\label{eq:16}
		\ket{\Phi}
			&=
				\sum_{r, R} \Phi(r, R)\ket{R+r / 2, R-r / 2}.
	\end{align}
$\Phi(r, R)$ is the exciton wavefunction and is given by the projection of the excited states on to the ground
state:
	\begin{align}\label{eq:17}
		 \Phi(r, R)
			&=
				\mel{\mathrm{GS}}{S_{rR}}{\Phi}.
	\end{align}
\par
\begin{figure}[h!]
				\centering
				\begin{subfigure}[b]{0.45\linewidth}
        					\includegraphics[width=\textwidth]{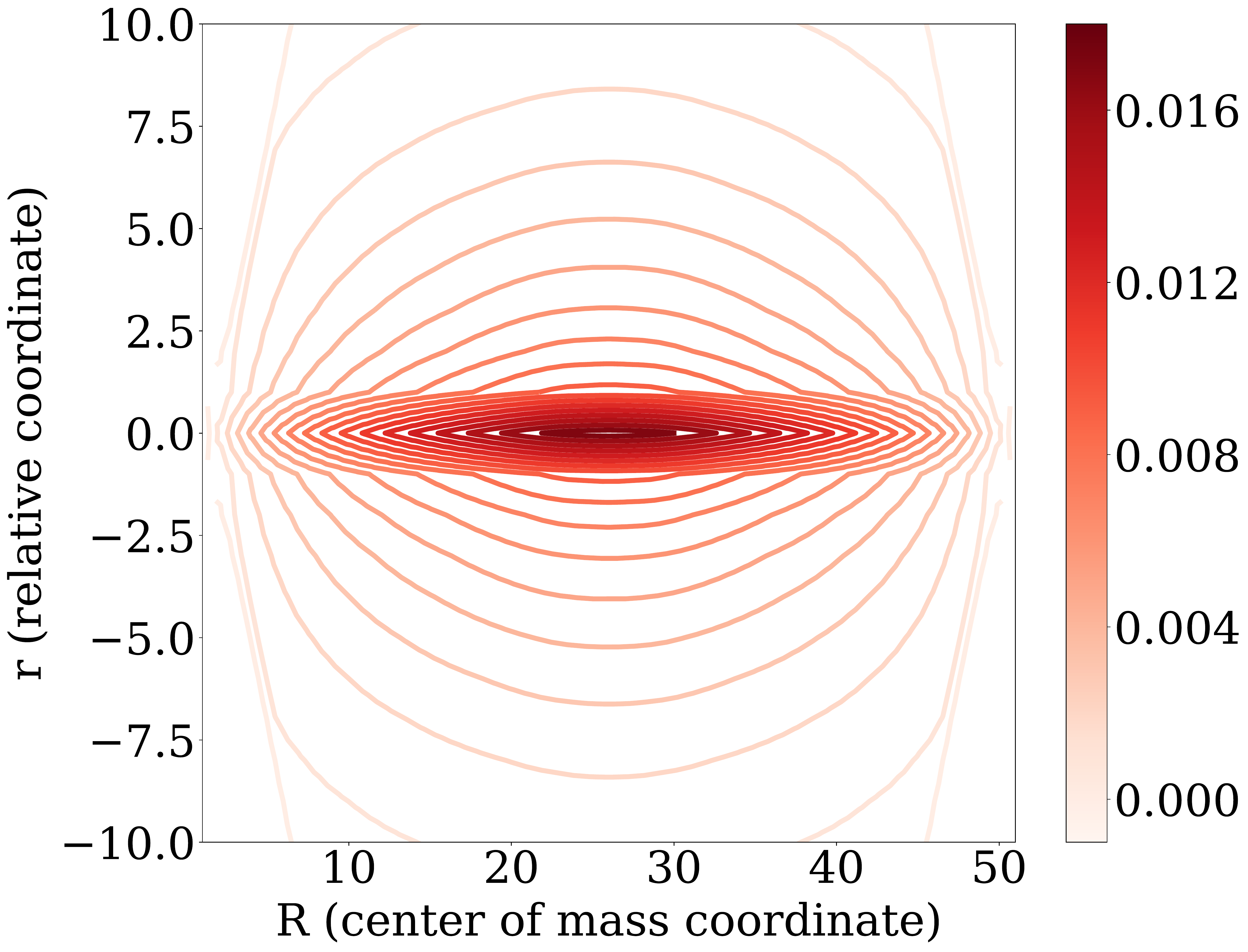}
					\caption{$1^1 B_u^+$ $(n=1, j=1)$}
    				\end{subfigure}
				\begin{subfigure}[b]{0.45\linewidth}
        					\includegraphics[width=\textwidth]{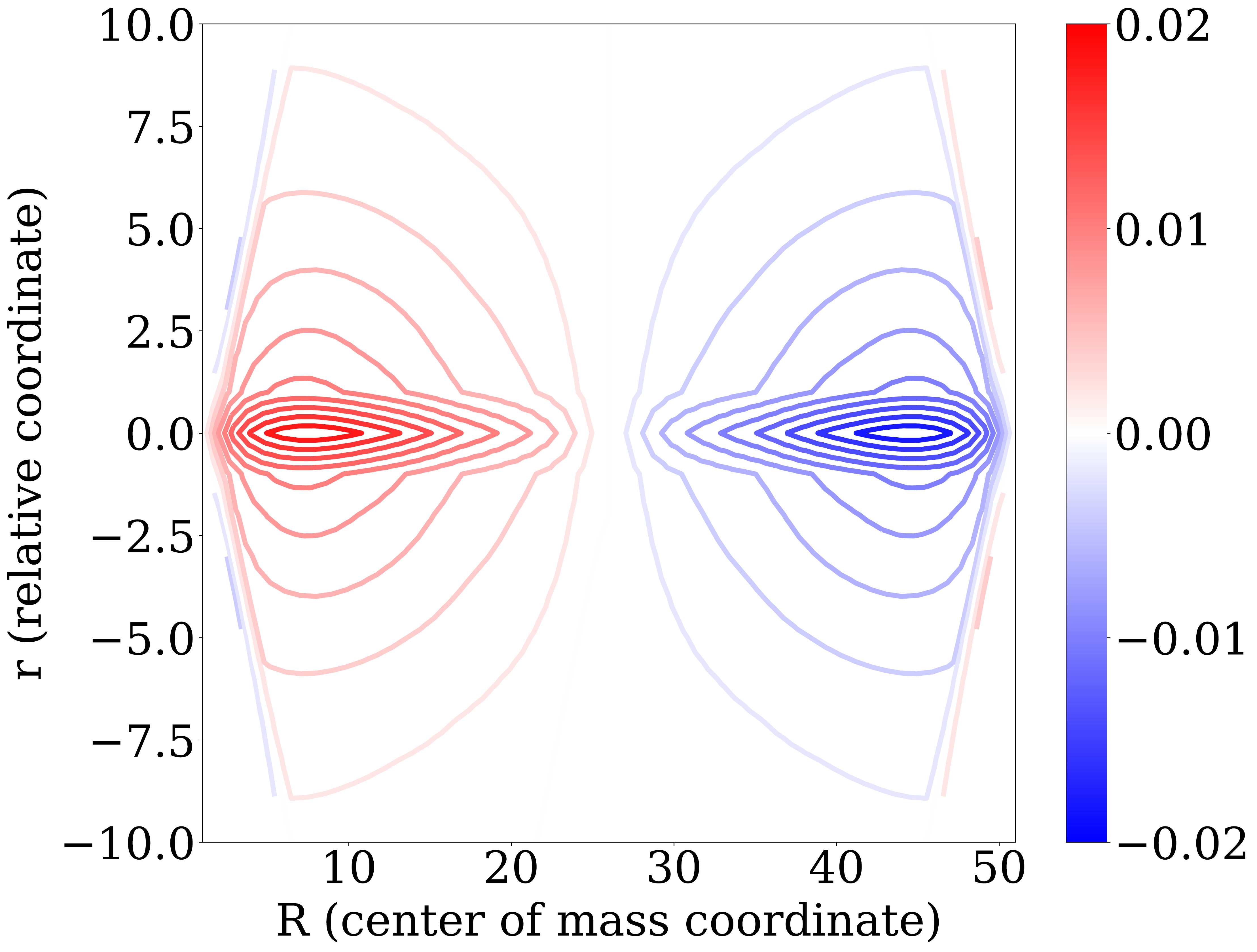}
					\caption{$1^1 A_g^+$ $(n=1, j=2)$}
    				\end{subfigure}
				\begin{subfigure}[b]{0.45\linewidth}
        					\includegraphics[width=\textwidth]{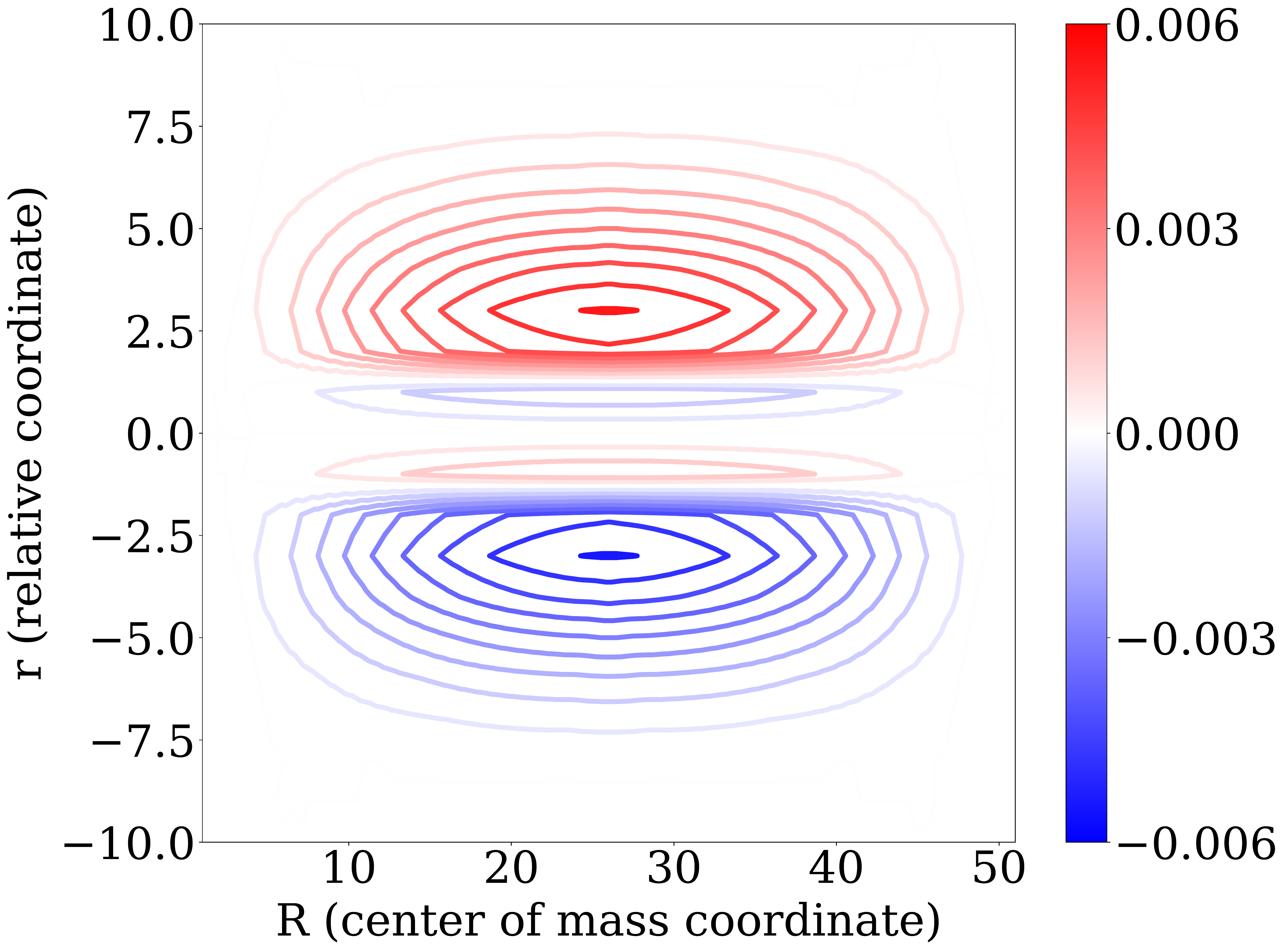}
					\caption{$2^1 A_g^-$ $(n=2, j=1)$}
    				\end{subfigure}
				\begin{subfigure}[b]{0.45\linewidth}
        					\includegraphics[width=\textwidth]{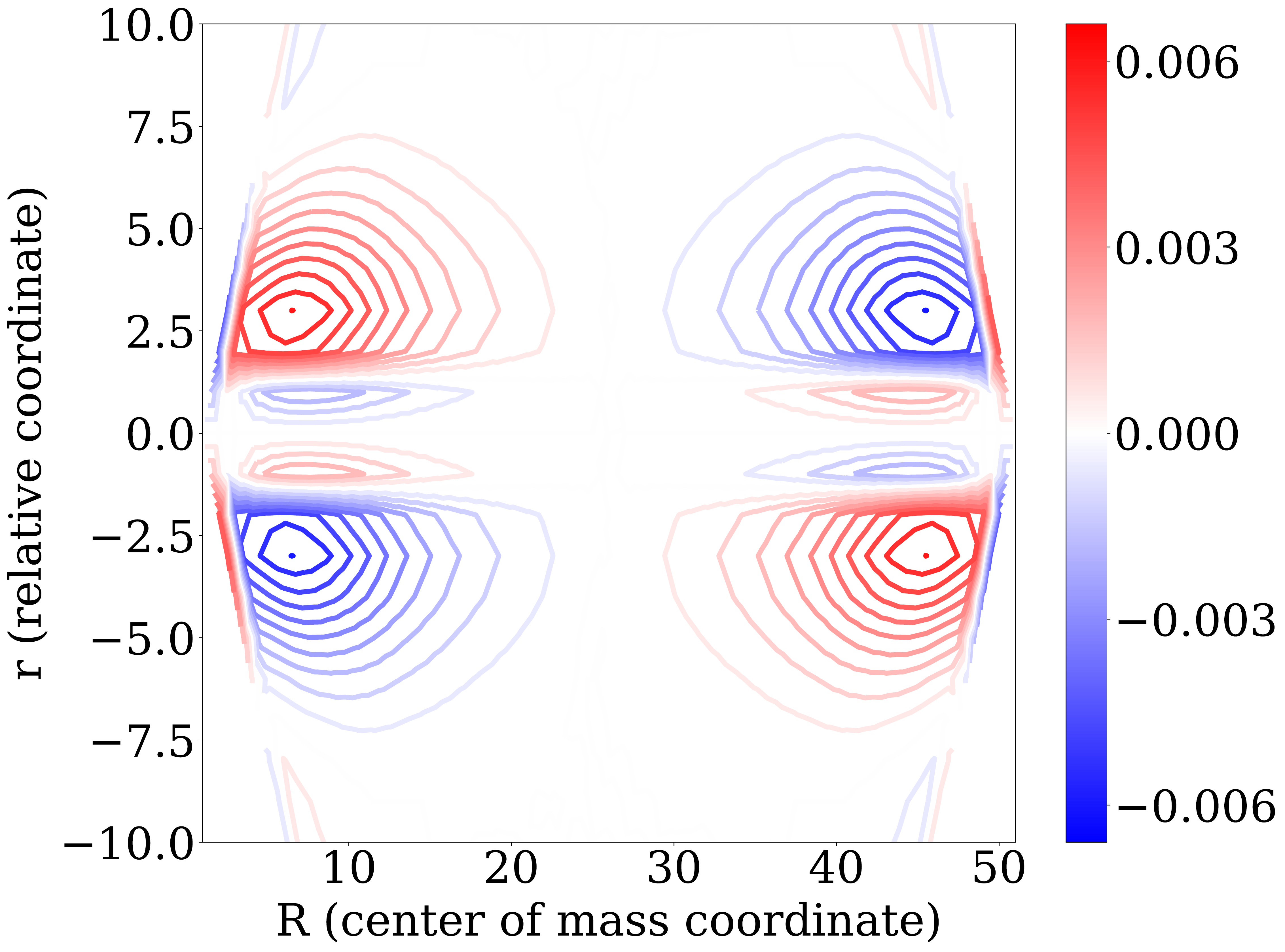}
					\caption{$1^1 B_u^-$ $(n=2, j=2)$}
    				\end{subfigure}
				\begin{subfigure}[b]{0.45\linewidth}
        					\includegraphics[width=\textwidth]{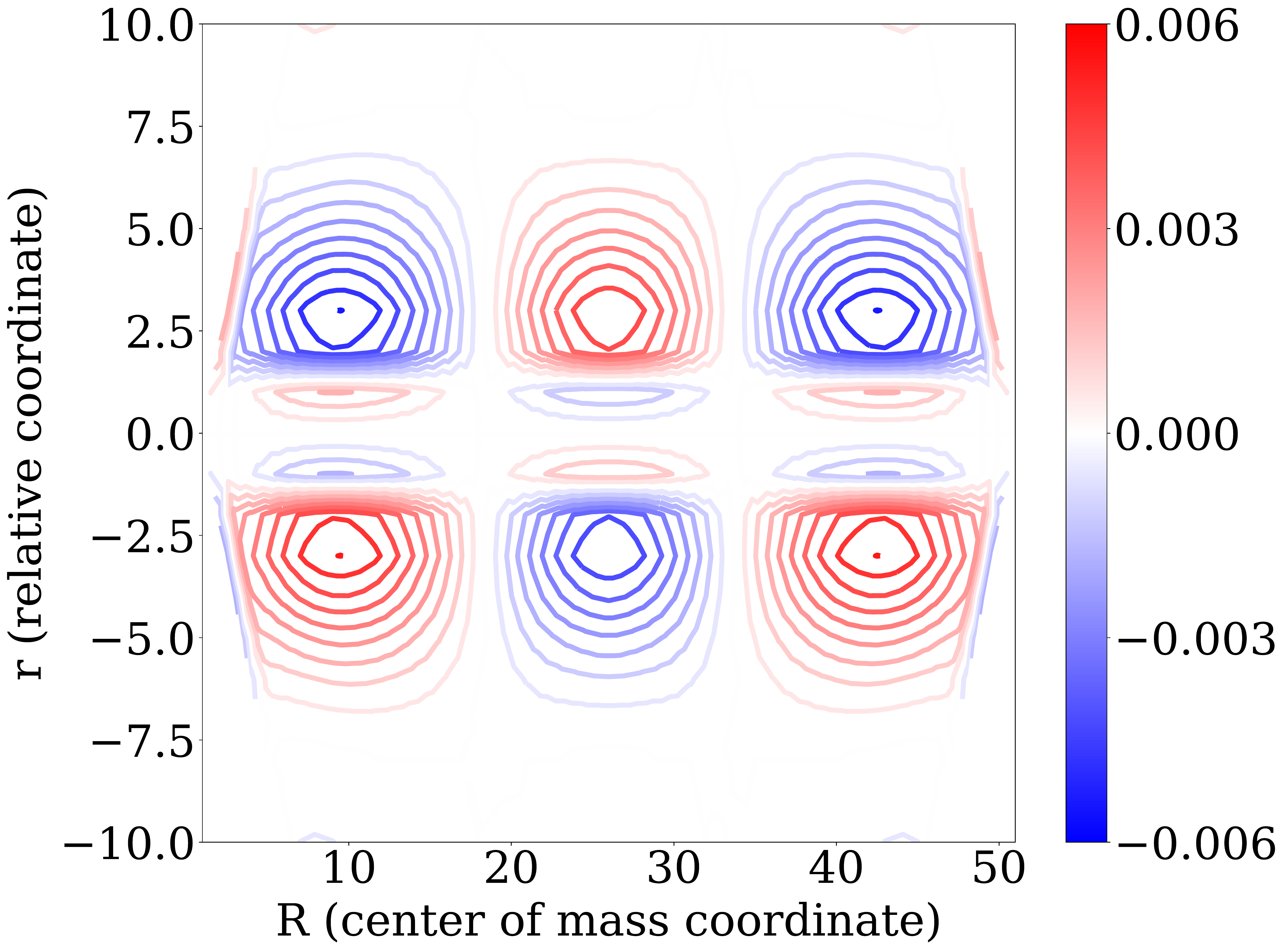}
					\caption{$3^1 A_g^-$ $(n=2, j=3)$}
    				\end{subfigure}
				\caption{Exciton components obtained from Eq.\ (\ref{eq:17}). $n$ and $j$ are the exciton principal and center-of-mass quantum numbers, respectively.}
				\label{exciton}
	\end{figure}

The calculated vertical exciton wavefunctions for the $1^1 B_u^+$,\ $1^1 A_g^+$,\ $2^1 A_g^-$,\ $1^1 B_u^-$ and
$3^1 A_g^-$ states of chain of $L=102$ are illustrated in Fig.\ \ref{exciton}.

\newpage
The nodal patterns of $\Phi (r,R)$ indicate that the $1^1B_u^+$ and $1^1A_g^+$ states have components belonging to the $n=1$ family of (even-parity) excitons with center-of-mass quantum numbers, $j$ = 1 and 2, respectively. Similarly, the $2^1A_g^-$, $1^1B_u^-$ and $3^1A_g^-$ states have components belonging to the $n=2$ family of (odd-parity) excitons with center-of-mass quantum numbers, $j$ = 1, 2 and 3, respectively. Thus, the single electron-hole components of the \agtwo, \agthree \ and \bum \ states belong to the same fundamental excitation. We note, however, that the electron-hole weights for these states are five times smaller than for the $1^1B_u^+$ and $1^1A_g^+$ states.

\subsection{The `$2A_g$ Family'}

As we have shown, the $2^1A_g^-$, $1^1B_u^-$ and $3^1A_g^-$ states have both triplet-pair and electron-hole components. For a translationally invariant system their vertical excitations can be expressed as
\begin{equation}\label{eq:65}
\ket{\Phi(K)} = a^{\textrm{TT}}(K)\ket{\Phi_{m}^{\textrm{TT}}(K)} +  a^{\textrm{e-h}}(K)\ket{\Phi_{n}^{\textrm{e-h}}(K)},
 \end{equation}
where $\ket{\Phi_{m}^{\textrm{TT}}(K)}$ is given by Eq.\ (\ref{eq:54}) and $\ket{\Phi_{m}^{\textrm{e-h}}(K)}$ is given by the Fourier transform of Eq.\ (\ref{eq:16}).
Both components are labelled by the same center-of-mass quantum number, $K$, but the principal quantum numbers for the triplet-pair ($m$) and electron-hole ($n$) components are different, being $1$ and $2$, respectively.

Equation (\ref{eq:65}) conveys the concept that the $2^1A_g^-$, $1^1B_u^-$ and $3^1A_g^-$ states are the three lowest-energy members of the same set of fundamental excitations which are distinguishable only by their center-of-mass momentum.

\section{Spectra}\label{sec:spec}

As the low-energy states have triplet-triplet components, we might expect their spectra to resemble that of the \triplet{} and $1^3A_g^-$  triplet states. We calculated the approximate spectra of the \agtwo, \agthree, \bum{}, \Q, \triplet{}($T_1$) and $1^3A_g^-(T_2)$  states for a $N=26$
using the expression,
\begin{equation}
	I(E) = \sum_i |\Braket{i|\hat{\mu}|\Psi}|^2 \delta(E_i -E_{\Psi} - E),
	\label{eqn:spectra}
\end{equation}
where the sum is over states with opposite particle-hole and $C_2$ symmetry to the state $\Ket{\Psi}$. $(E_i - E_{\Psi})$ is the energy difference between state $i$ and state $\Psi$, $\hat{\mu}$ is the transition dipole operator and $ |\Braket{i|\hat{\mu}|\Psi}|^2$ is square of the transition dipole moment between states $i$ and $\Psi$.

\begin{figure*}
	\centering
	\begin{subfigure}{\linewidth}
		\xincludegraphics[width=0.8\linewidth, label=(a)]{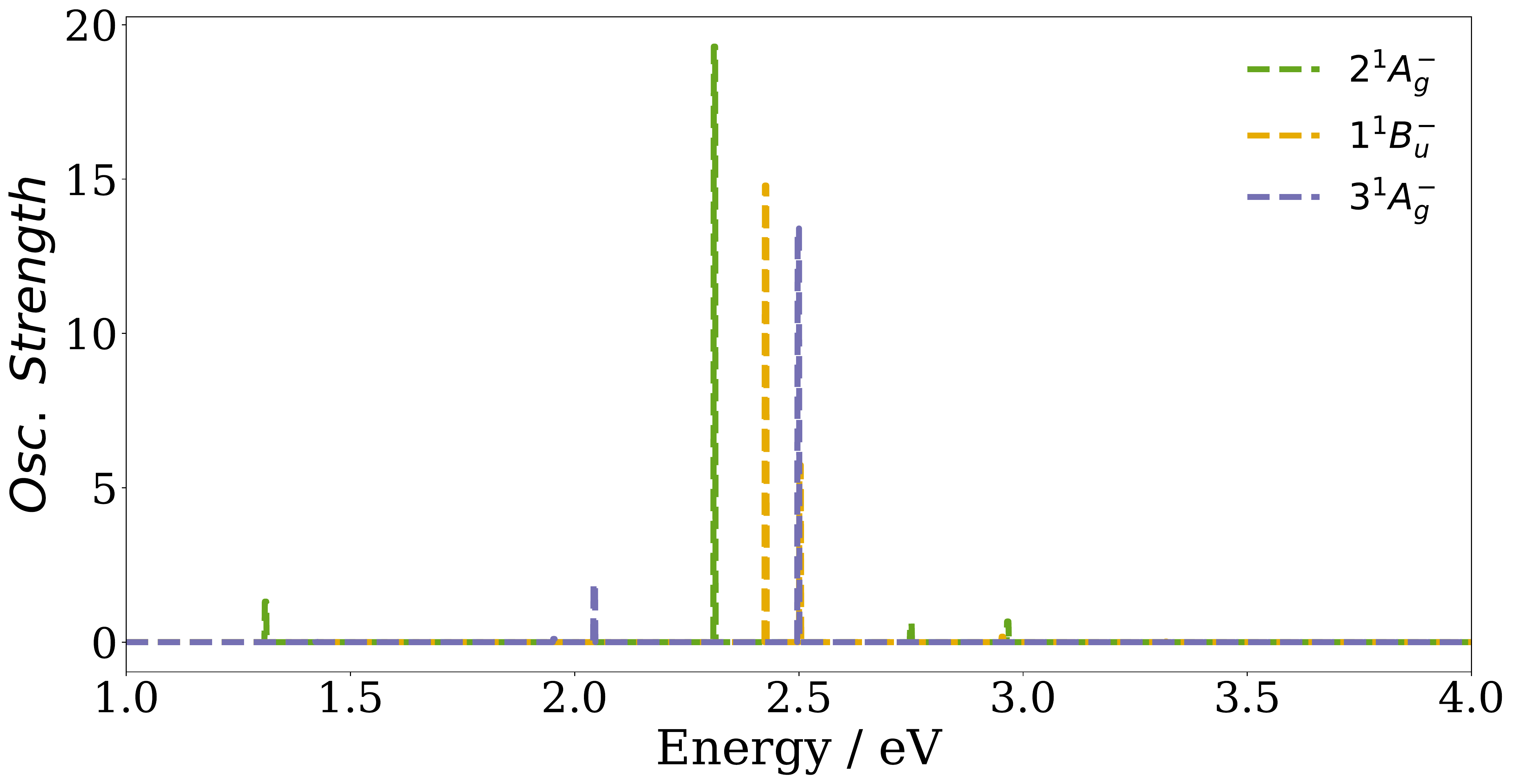}
	\end{subfigure}
	\begin{subfigure}{\linewidth}
		\xincludegraphics[width=0.8\linewidth, label=(b)]{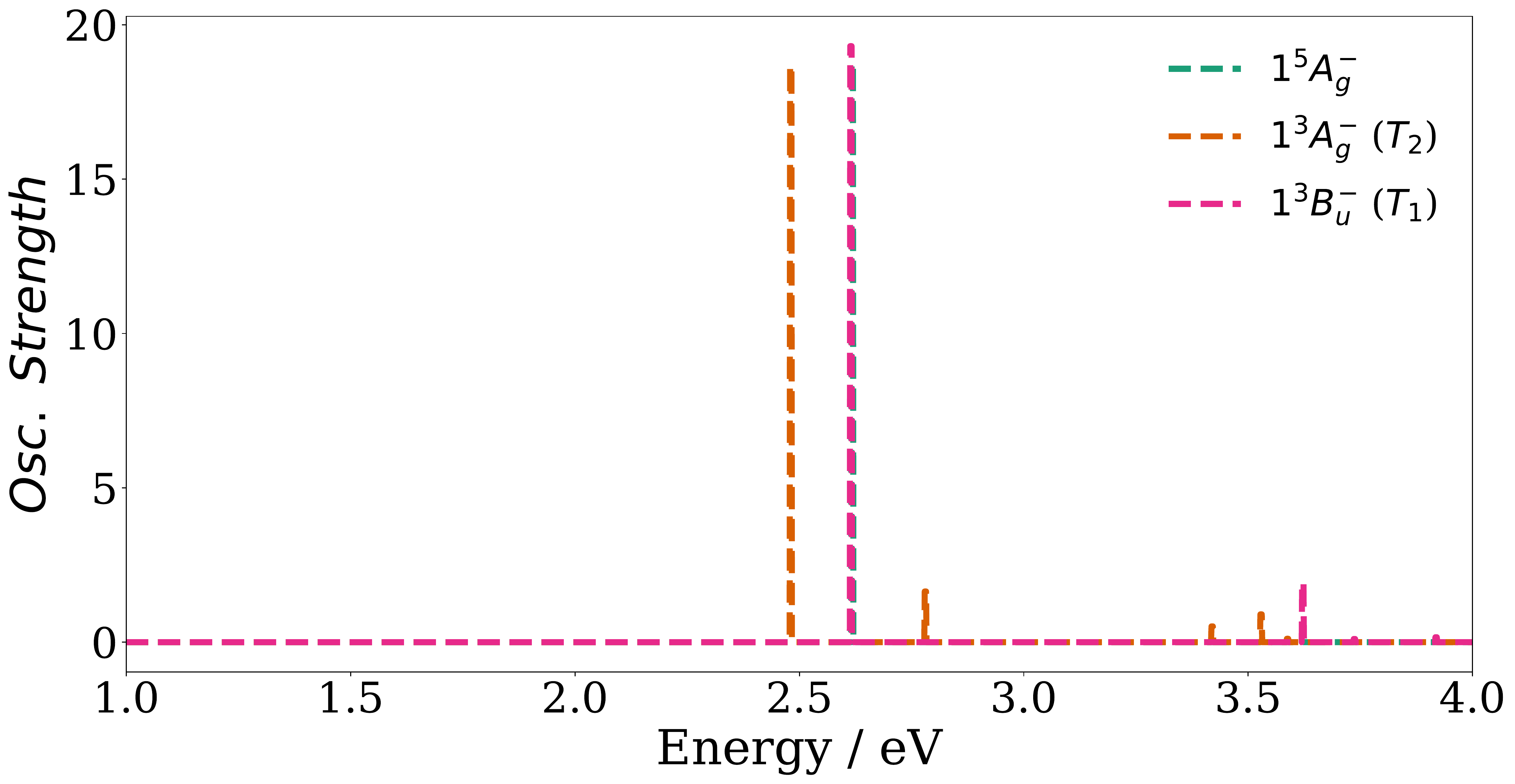}
	\end{subfigure}
	\caption{Stick spectra  calculated using Eq.\ \ref{eqn:spectra}  of the  (a) \agtwo, \agthree and \bum \  states and (b) \Q, \triplet{}($T_1$) and $1^3A_g^-$ $(T_2)$  for 26 site chain}
	\label{fig:spectra_26}
\end{figure*}

\begin{figure*}
	\centering
	\begin{subfigure}{\linewidth}
		\xincludegraphics[width=0.8\linewidth, label=(a)]{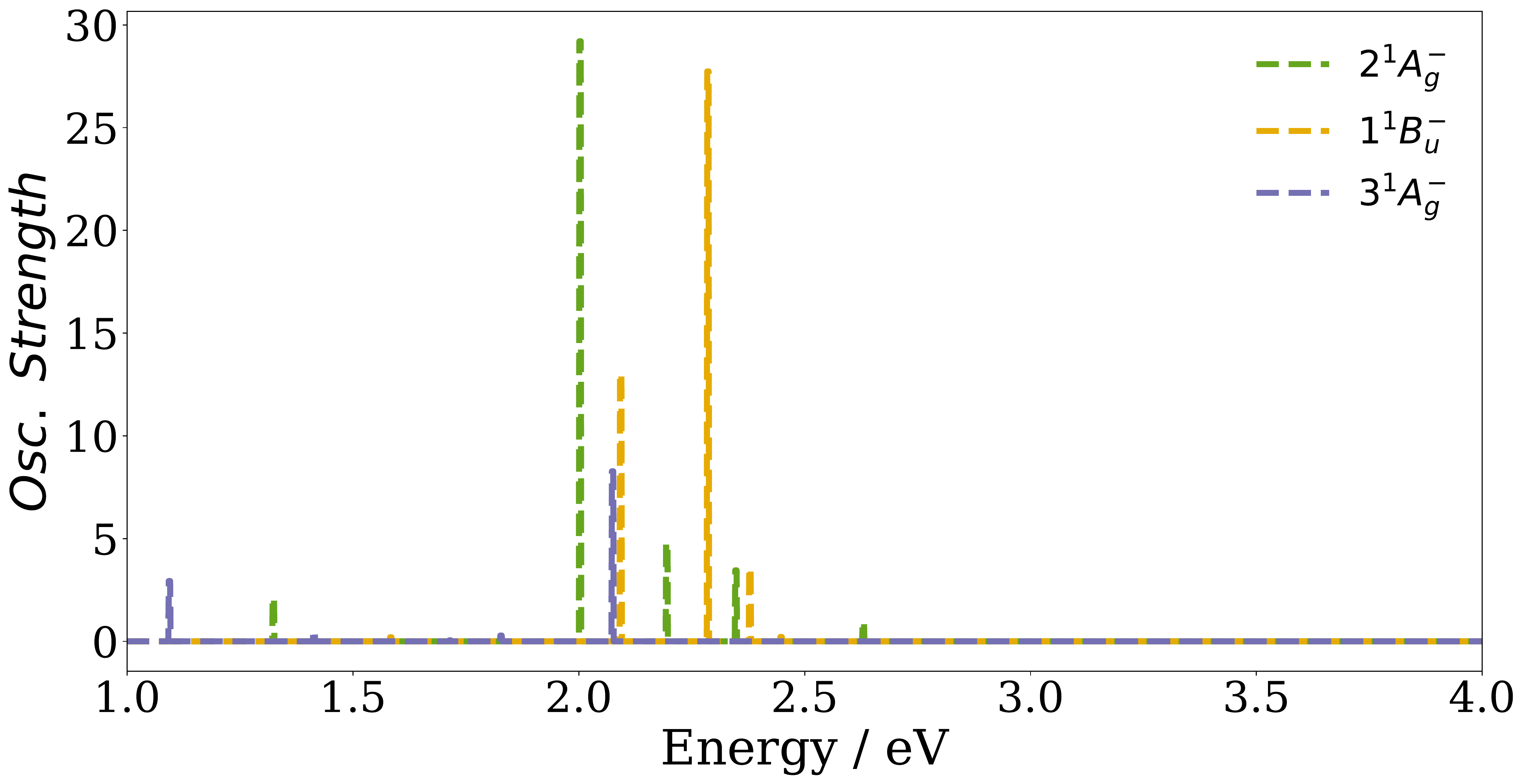}
	\end{subfigure}
	\begin{subfigure}{\linewidth}
		\xincludegraphics[width=0.8\linewidth, label=(b)]{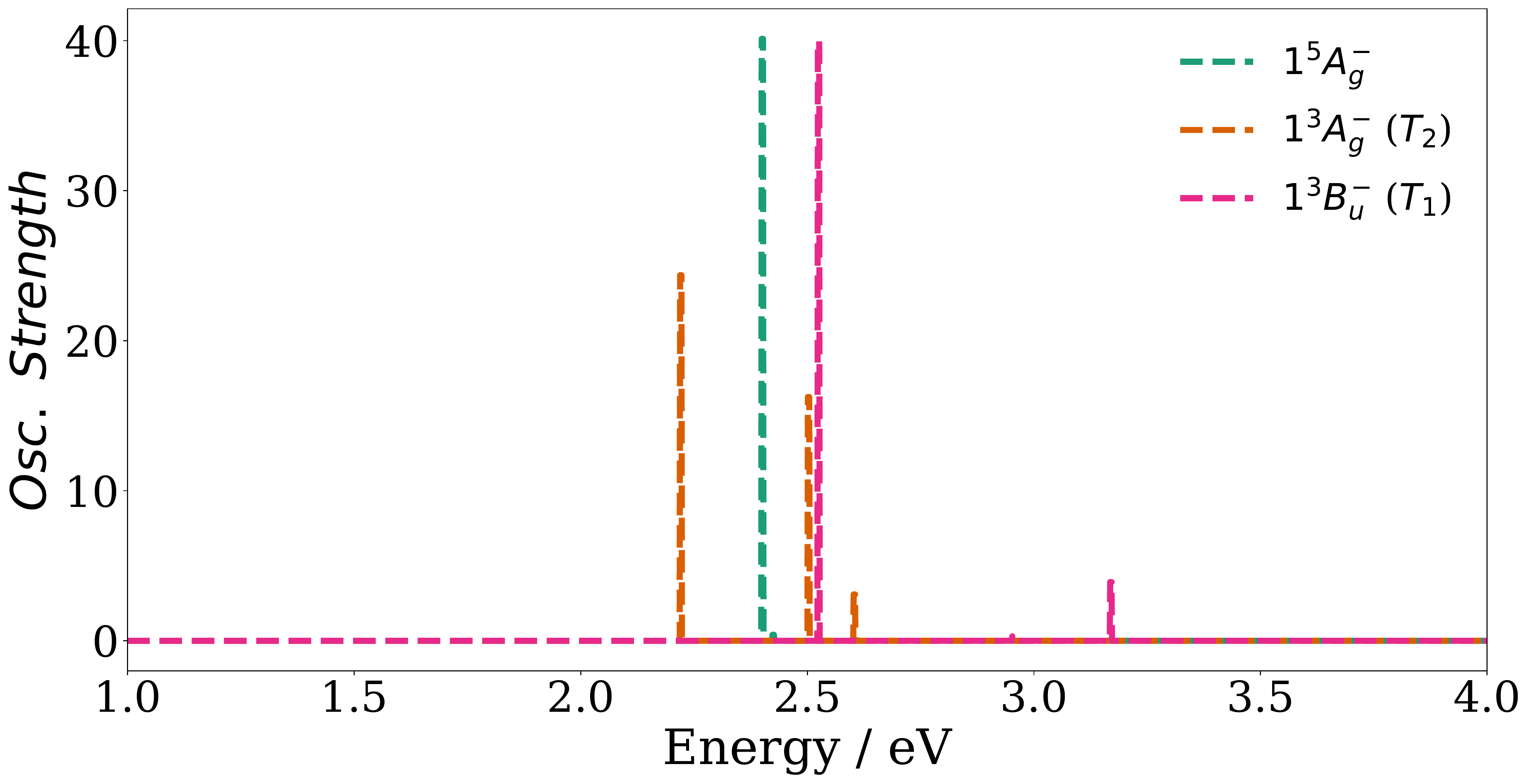}
	\end{subfigure}
	\caption{Stick spectra  calculated using Eq.\ \ref{eqn:spectra}  of the  (a) \agtwo, \agthree and \bum \  states and (b) \Q, \triplet{}($T_1$) and $1^3A_g^-$ $(T_2)$  for 54 site chain}
	\label{fig:spectra_54}
\end{figure*}

As shown in
Figs. \ref{fig:spectra_26} -\ref{fig:spectra_54},
for all of the triplet-pair states, the maximum absorption occurs  within $\sim 0.5$ eV of the \triplet \ maximum absorption. Given that the \agtwo \  state is considered to be a bound triplet-pair, we might expect that the maximum absorption energy to be higher than the triplet state, as a photoexcitation would need to overcome the binding energy of the triplet, in addition to having enough energy to excite a triplet state. However, the maximum absorption of the \agtwo \ state is found to be lower than the \triplet \ state, in agreement with experimental observations in carotenoids \cite{Polak2019}.  The \agtwo \ state also has an absorption in the near infra-red part of the spectrum, which can be attributed to the \agtwo $\rightarrow$ \bu \ transition. The near infra-red absorption of a bound triplet-pair has also been predicted in acene materials \cite{Khan2020}.

The \Q{} quintet state exhibits a single absorption with energy closest to the triplet maximum absorption over all chain lengths and whose intensity most closely matches the  \triplet{} absorption. Comparing the \bum{}  to the triplets, for each triplet absorption there is a corresponding red-shifted absorption in the \bum{} spectra, also indicating that despite being a bound state, absorption are lower in energy compared to individual triplets.

Due to the mixed triplet-pair character of the \agthree{} state there are many different absorptions. As for the \agtwo state, the \agthree{} state has a lower energy absorption in the infra-red to yellow portion of the spectrum.

We note that although our calculated transition energy from the \Q{} state coincides with our calculated $T-T^*$ transition energy (i.e., ca. 2.56 eV  for the 26-site chain) this energy is over an eV  higher than the observed $T-T^*$ transition energy for singlet fission in conjugated polyenes \cite{Musser2013, Musser2019}. We explain this discrepancy to the failure of the Mazumdar and Chandross parameterization of the PPP model to correctly estimate the solvation energy of weakly bound excitons and charges \cite{Chandross1997}. The $T^*$ state is expected to be the $n=2$ (charge-transfer) triplet exciton, whose solvation energy is over an eV larger than predicted by the parametrized PPP model \cite{Barford2011}.

\section{Discussion and Conclusions}\label{sec:con}

By calculating the relaxed energies of the singlet states of conjugated polyenes, we find that the  \bum \   and \agthree \ states lie below the bright \bu \  state at experimentally relevant chain lengths. This implies that these states could be involved in relaxation pathways, particularly if systems are excited with energy higher than the band edge. In addition, we find that the energy of the relaxed \Q{} state on a chain of $N$ C-atoms is twice the energy of the relaxed triplet state on a chain of $N/2$ C-atoms, so if spin mixing where allowed this state could represent an intermediate unbound triplet-pair state for the singlet fission process.

An analysis of the bond dimerization of the relaxed excitations indicates that the \agtwo \ is a four-soliton state, as previously found \cite{Hayden1986}. The \Q{}  \ and \bum \ states are also found to be four-soliton states. Both of these states seem to consist of repelling soliton pairs, with the bond dimerization of the \Q{} \ resembling two \triplet \ triplets occupying either side of the chain. The \agthree state bond dimerization is more complicated due to the mixed triplet-pair combinations that contribute to this state.
	
The spin-spin correlation function offers another way to visualise the soliton structure. This again indicates that the \agtwo \ is a bound triplet-pair. We also find that the \Q{} and  \bum{} states show long-range spin correlations, which correspond to the staggered bond dimerisation.

The calculated spectra indicate that the \Q{} \ state most closely resembles the triplet absorption, although the \bum \ and \agthree \ states also absorb at a similar energy.
Recent pump-push-probe experiments by Pandya et al.\ excited the \agtwo \ state (push) after being generated from the relaxation of the initially photoexcited state \cite{Pandya2020}. As the \agtwo \ state has $^1(T_1T_1)$ character, the excited push state is expected to be of $^1(T_1T^*)$ character. Relaxation from this state was found to involve a state with spatially separated, but correlated triplet pairs. We predict that this state is either the  \bum{} or \agthree{} state \cite{Pandya2020}.

We further probed the triplet-triplet nature of the \agtwo, \agthree \ and \bum \ states by calculating the overlap of these states with half-chain triplet combinations. The $ T_1 \otimes T_1 $ nature of the \agtwo \ and $ T_1 \otimes T_2 $ nature of the \bum \ state was confirmed. The \agthree \ state has a mixture of $ T_1 \otimes  T_1 $, and symmetry allowed $T_1 \otimes T_2$ and $ T_2 \otimes T_2 $ contributions.  We also showed that the electron-hole excitation components of the \agtwo, \bum \ and \agthree \ states belong to the same $n=2$ family of excitons with center-of-mass quantum numbers $j=1,2,$ and 3, respectively.

One of the aims of this work has been to identify a singlet state in polyenes that is intermediate between the initially photoexcited singlet state, $S_2$, and the final non-geminate pair of triplet states.
Such a state should satisfy the following conditions:
\begin{enumerate}
\item{It should have significant triplet-triplet character.}
\item{Its vertical energy should lie above the vertical energy of $S_2$, but its relaxed energy should lie below the relaxed energy of $S_2$. Such conditions imply the possibility of an efficient interconversion from $S_2$ via a conical intersection.}
\item{Its relaxed energy should lie slightly higher than twice the relaxed energy of the triplet state, so that fission is fast and exothermic.}
\end{enumerate}
For our choice of model parameters we find that:
the $2^1A_g^-$ state only satisfies condition (1.); the $1^1B_u^-$ state satisfies condition (1.), conditions (2.) for $10 < N < 26$, but not condition (3.); the $3^1A_g^-$ state satisfies condition (1.), conditions (2.) for $20 < N < 46$, and condition (3.). Thus, the $3^1A_g^-$ state would appear to be a candidate intermediate state for longer polyenes, but such a state does not exist for shorter carotenoids.

We should be cautious, however, about making a prediction about the precise intermediate state, as owing to using semi-empirical parameters our calculated excitation energies are only expected to be accurate to within a few tenths of an eV. Our key conclusion, therefore, is that there is a family of singlet excitations (the `\agtwo family'), composed of both triplet-pair and electron-hole character, which are fundamentally the same excitation (i.e., have the same principal quantum numbers), but have different center-of-mass energies. The lowest energy member of this family, the $2^1A_g^-$ state, cannot undergo singlet fission. But higher energy members, owing to their increased kinetic energy and reduced electron-lattice relaxation, can undergo singlet fission for certain chain lengths.

We are currently investigating the dynamics of interconversion to the `\agtwo family'  from $S_2$ using time-dependent DMRG.
It is tempting to assign the $3^1A_g^-$ state (or one of its relatives) as  the geminate triplet-pair, often denoted as $^1(T \cdots T)$. A possible mechanism to explain how this state undergoes spin decoherence to become a non-geminate pair is described in the recent paper by Marcus and Barford \cite{Marcus2020}.
	
\begin{acknowledgments}
The authors thank Jenny Clark and Max Marcus for helpful discussions. D.V and D.M. would like to thank the EPSRC Centre for Doctoral Training, Theory and Modelling in Chemical Sciences, under Grant No. EP/L015722/1, for financial support. D.V. would also like to thank Balliol College Oxford for a Foley-B\'{e}jar Scholarship. D.M. would also like to thank  Linacre College for a Carolyn and Franco Giantruco Scholarship and the Department of Chemistry, University of Oxford.
\end{acknowledgments}

\bibliography{library}

\end{document}